\documentclass[]{IEEEtran}

\usepackage{cite}

\ifCLASSINFOpdf

\else

\fi

\usepackage[switch, pagewise]{lineno}

\usepackage{mathtools}

\usepackage{graphicx}
\PassOptionsToPackage{hyphens}{url}\usepackage[hidelinks]{hyperref}
\usepackage[utf8]{inputenc}
\usepackage[english]{babel}
\usepackage{amsthm}
\usepackage{amsmath}
\usepackage{MnSymbol}
\usepackage{wasysym}%
\usepackage[short]{optidef}
\usepackage{makecell}

\DeclareFontFamily{U}{mathx}{\hyphenchar\font45}
\DeclareFontShape{U}{mathx}{m}{n}{
      <5> <6> <7> <8> <9> <10>
      <10.95> <12> <14.4> <17.28> <20.74> <24.88>
      mathx10
      }{}
\DeclareSymbolFont{mathx}{U}{mathx}{m}{n}
\DeclareFontSubstitution{U}{mathx}{m}{n}
\DeclareMathAccent{\widecheck}{0}{mathx}{"71}
\DeclareMathAccent{\wideparen}{0}{mathx}{"75}

\usepackage{amsfonts}
\usepackage{chngcntr}
\usepackage{apptools}
\usepackage{pgfgantt}
\usepackage{lscape}
\usepackage[pass]{geometry}
\usepackage{booktabs}
\usepackage{algpseudocode} 
\ifCLASSOPTIONcompsoc
    \usepackage[caption=false, font=normalsize, labelfont=sf, textfont=sf, subrefformat=parens,labelformat=parens]{subfig}
\else
\usepackage[caption=false, font=footnotesize, subrefformat=parens,labelformat=parens]{subfig}

\usepackage{algorithm}
\usepackage{array}
\usepackage{tabularray}
\usepackage{afterpage}

\AtAppendix{\counterwithin{lemma}{section}}
\theoremstyle{definition}

\theoremstyle{plain}
\usepackage{multirow}

\usepackage{xcolor}
\usepackage{cases} 
\usepackage{empheq}
\usepackage{enumitem}
\usepackage{graphicx}

\hypersetup{
  colorlinks=true,
  linkcolor=blue,
  citecolor=blue,
  urlcolor=black
}

\DeclareRobustCommand*{\IEEEauthorrefmark}[1]{%
  \raisebox{0pt}[0pt][0pt]{\textsuperscript{\footnotesize #1}}%
}

\usepackage{nomencl}
\makenomenclature

\usepackage{etoolbox}
\renewcommand\nomgroup[1]{%
  \item[\bfseries
  \ifstrequal{#1}{P}{Parameters}{%
  \ifstrequal{#1}{N}{Number sets}{%
  \ifstrequal{#1}{A}{Acronyms}{%
  \ifstrequal{#1}{F}{Functions}{%
  \ifstrequal{#1}{X}{Decision variables}{}}}}}%
]}

\begin{document}
\bstctlcite{IEEEexample:BSTcontrol}

\title{AI-focused HPC Data Centers Can Provide More Power Grid Flexibility and at Lower Cost}

\newif\ifdoubleblind
\doubleblindfalse 

\author{
\ifdoubleblind
    Anonymous Authors\\
    Anonymous Affiliation
\else
    \IEEEauthorblockN{Yihong Zhou\IEEEauthorrefmark{1},
    \'Angel Paredes\IEEEauthorrefmark{2},
    Chaimaa Essayeh\IEEEauthorrefmark{3},
    and Thomas Morstyn\IEEEauthorrefmark{4}}
    \\
    \vspace{2mm}
    \IEEEauthorblockA{\IEEEauthorrefmark{1}School of Engineering, The University of Edinburgh, U.K., yihong.zhou@ed.ac.uk}\\
    \IEEEauthorblockA{\IEEEauthorrefmark{2}Department of Electrical Engineering, University of M\'alaga, Spain, angelparedes@uma.es} \\
    \IEEEauthorblockA{\IEEEauthorrefmark{3}Department of Engineering, Nottingham Trent University, U.K., chaimaa.essayeh@ntu.ac.uk}\\
    \IEEEauthorblockA{\IEEEauthorrefmark{4}Department of Engineering Science, University of Oxford, U.K., thomas.morstyn@eng.ox.ac.uk}\\
    \vspace{-0 mm}
\fi
}

\maketitle
\begin{abstract}
The recent growth of Artificial Intelligence (AI), particularly large language models, requires energy-demanding high-performance computing (HPC) data centers, which poses a significant burden on power system capacity. Scheduling data center computing jobs to manage power demand can alleviate network stress with minimal infrastructure investment and contribute to fast time-scale power system balancing. This study, for the first time, comprehensively analyzes the capability and cost of grid flexibility provision by GPU-heavy AI-focused HPC data centers, along with a comparison with CPU-heavy general-purpose HPC data centers traditionally used for scientific computing. A data center flexibility cost model is proposed that accounts for the value of computing. Using real-world computing traces from 7 AI-focused HPC data centers and 7 general-purpose HPC data centers, along with computing prices from 3 cloud platforms, we find that AI-focused HPC data centers can offer greater flexibility at 50\% lower cost compared to general-purpose HPC data centers for a range of power system services.
By comparing the cost to flexibility market prices, we illustrate the financial profitability of flexibility provision for AI-focused HPC data centers. Finally, our flexibility and cost estimates can be scaled using parameters of other data centers through algebraic operations, avoiding the need for re-optimization. 
\end{abstract}


\IEEEpeerreviewmaketitle

\section{Introduction}

The rapid development of Artificial Intelligence (AI) has attracted significant interest from researchers, industry, policy makers, and the general public. The most representative example is Large Language Models (LLMs) with versatile language understanding, such as ChatGPT. AI is also widely applied in other domains, such as computer vision and health care \cite{Demszky2023, Singhal2023}. To support rapid development and widespread use, there are now high-performance computing (HPC) data centers dedicated to AI \cite{seren}. To fully exploit the parallelizability of AI jobs, these AI-focused HPC data centers are accelerator-intensive \cite{seren, runai_HPC_cpu_gpu}, and the most typical accelerator is the Graphics Processing Unit (GPU). Our paper uses the term ``GPU", but it should be noted that other types of accelerator exist, such as the Tensor Processing Unit (TPU) in Google. 

Prior to the emergence of AI-focused HPC data centers, general-purpose HPC data centers were used for scientific computing applications, such as simulating complex physical systems and solving large-scale mathematical programming problems. Large-scale general-purpose HPC data centers include those at Oak Ridge National Laboratory (ORNL) \cite{ornlLandingPage} and Argonne Leadership Computing Facility (ALCF) \cite{anlALCFPublic}. While GPUs are present in these general-purpose HPC data centers, they primarily use Central Processing Unit (CPUs) and many of their jobs are not amenable to GPU acceleration \cite{vestias2014trends, runai_HPC_cpu_gpu}. 

AI-focused HPC data centers can be more energy-demanding than general-purpose HPC data centers due to their heavy use of energy-demanding GPUs. For example, a latest-generation ``192-thread AMD EPYC 9654 CPU'' has a rated power of only 360 W, while the latest-generation ``NVIDIA B200 GPU'' can draw 1000 W. The other contributor is the rapidly expanding use of LLMs, which requires intensive computing \cite{Carbon-Emissions-and-Large-Neural-Network-Training}. Venture Capital firms invested \$290 billion in AI over the last 5 years \cite{venture_capital}, suggesting substantial growth in the field. Estimated by Electric Power Research Institute (EPRI), data centers now consume 4\% of the total power generation annually in the US, and this proportion can increase to as much as 9.1\% by 2030, with AI being one of the main drivers \cite{US_dace_demand}. The increased energy demand for AI poses a major challenge for power grid infrastructure \cite{Mytton2023}, especially when considered alongside the ongoing efforts to electrify heating and transportation to achieve net-zero carbon emissions \cite{zhou2024evaluating}.
Being aware of this challenge, AI companies and hardware manufacturers are making their systems more energy efficient \cite{AI_carbon_decline_2030, nvidia_blackwell}. However, Jevons Paradox suggests that increasing energy efficiency may lead to greater demand for computing and higher energy usage \cite{Jevons_para}. 

Another avenue is to make data center power demand flexible, that is, adjusting a data center's computing workload and consequently its power demand. This flexibility can be used for power system services that are critical to maintain normal and stable power system operation \cite{paredes2023stacking}. For example, Refs. \cite{Alapera2018} and \cite{anc_datacenter} studied the data center flexibility in maintaining power system frequency. In \cite{Chen2021}, the data center flexibility was used for mitigating voltage and imbalance issues in distribution networks. Ref. \cite{Zhang2022} explored the data center flexibility in regulation services, where data centers track power system signals every few seconds. Ref. \cite{cao2024managing} further studied the use of data center flexibility to participate in energy balancing market. These studies investigated data center flexibility in a wide range of power system services. However, there is a lack of assessment of these diverse power system services within one study, making it difficult to understand the capability of the same data center in providing different power system services. Moreover, these studies did not distinguish between AI-focused and general-purpose HPC data centers. As mentioned, AI-focused HPC data centers can be more power-demanding due to the heavy use of GPUs, which may lead to different capabilities in providing power system services. In addition, recent work \cite{GPU_dataset} observed that there exist distinct job patterns in the two types of data centers, which can also affect the capability in power system services. Finally, although some work imposed a limit on the disruption of data center computing workload \cite{Zhang2022, anc_datacenter, ding_datacenter_rl}, this disruption was not quantified as a cost. Data center operators may still remain disincentivized due to the unclear financial profitability of providing power system services by (potentially) disrupting valuable computing jobs.

To fill these gaps, this paper evaluates the maximum power flexibility of AI-focused HPC data centers compared to general-purpose HPC data centers for a comprehensive set of power system services. In addition, we propose a method for estimating the cost incurred by data centers providing power system services. This cost estimation accounts for the value of computing by using data from three cloud computing platforms (Google Cloud, Amazon AWS, and Oracle). Our analysis uses real-world datasets of 7 AI-focused HPC data centers and 7 general-purpose HPC data centers. These datasets vary between 40 to 80 days, with a temporal resolution of one second or even finer, and have between 13,397 and 962,602 computing jobs. Table \ref{tab:summary} (at the end of the paper) provides a detailed description of these datasets.
We find that AI-focused HPC data centers can provide greater power flexibility and at 50\% lower cost than general-purpose HPC data centers for a range of power system services, which enables AI-focused data centers to be more competitive for the same market conditions. By comparing flexibility cost and real-world power system service prices, we illustrate the financial profitability of flexibility provision for AI-focused HPC data centers. This may break the stereotype that computing jobs are always more valuable than providing power system services, and brings financial motivations for data center operators to provide power system services. 
Additionally, a correlation analysis illustrates that data center utilization patterns also contribute to the greater flexibility and lower cost of AI-focused HPC data centers. Finally, we investigate the opportunity for dynamic quotas for parallelizable jobs to increase data center flexibility and reduce flexibility provision cost. The superiority of AI-focused HPC data centers still persists in this opportunity. 

An advantage of our novel methodology is that our flexibility and cost estimates can be scaled using parameters of other data centers through algebraic operations, avoiding the need for re-optimization. These scaling formulas can be found in the \nameref{sec:method} section. We have developed an interactive Google Colab page \cite{dace_api} which implements this functionality, making our analysis approach accessible to anyone to have quick assessment with their own data center information.




\section{Results}
\label{sec:results}

\begin{figure}[tb]
    \centering
    \subfloat[ ]{\includegraphics[width=1\linewidth]{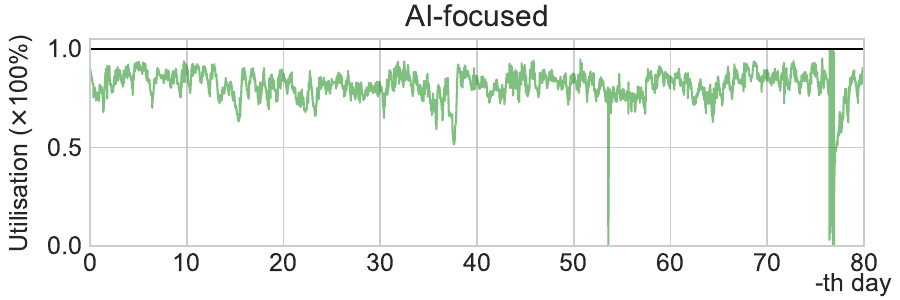}\label{fig:uti_base_plot_AI}}
    
    \subfloat[ ]{\includegraphics[width=1\linewidth]{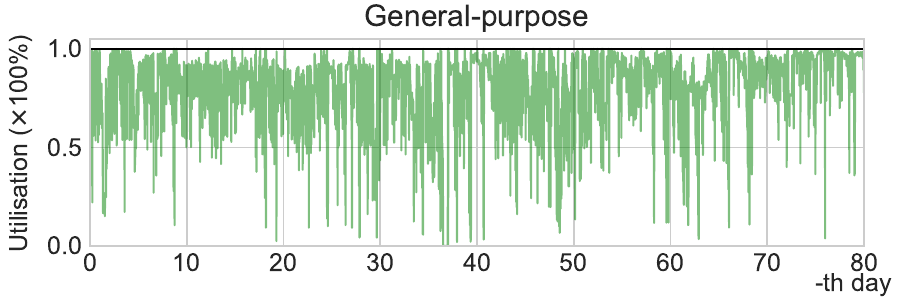}\label{fig:nuti_base_plot_General}}
    
    \subfloat[ ]{\includegraphics[width=1\linewidth]{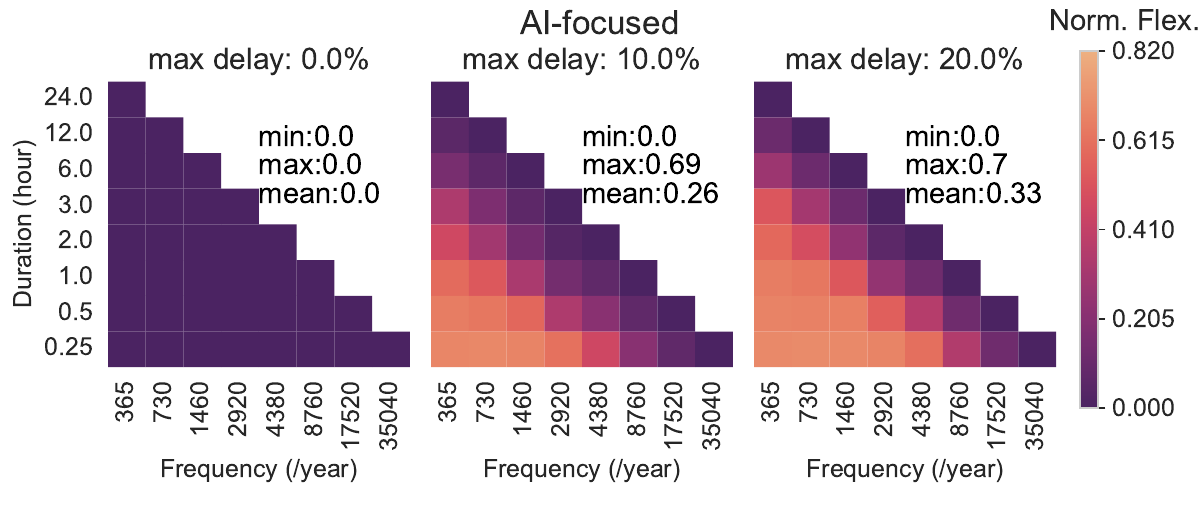}\label{fig:normalised_flex_AI}}
    
   \subfloat[ ]{\includegraphics[width=1\linewidth]{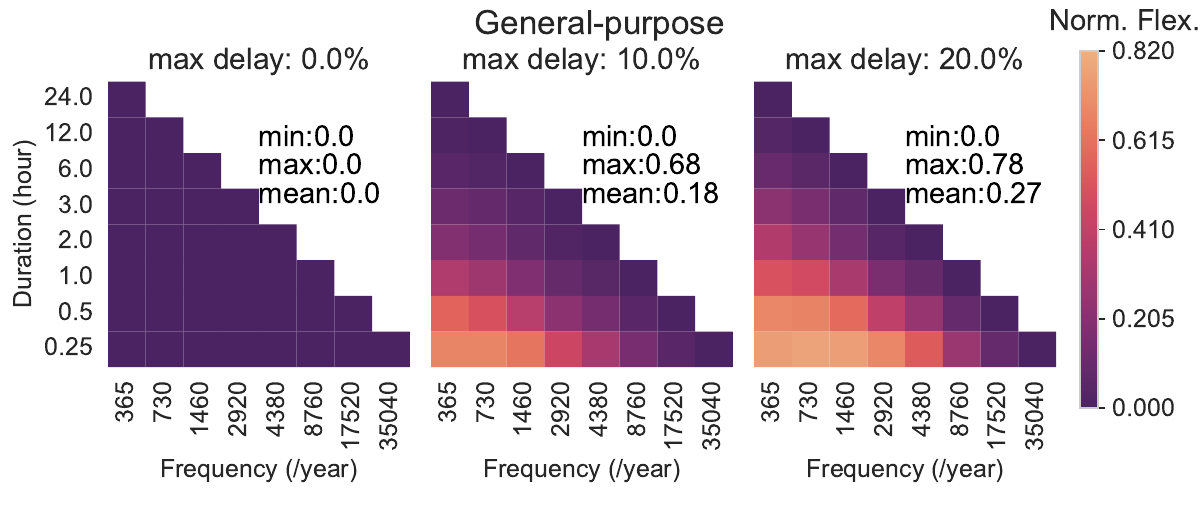}\label{fig:normalised_flex_General}}
   \vspace{-6 mm}
    \caption{Baseline utilization time series and maximum amount of flexibility for two selected data centers. (a) Utilization of the AI-focused HPC data center (Saturn in Table \ref{tab:summary}). (b) Utilization of the general-purpose HPC data center (ORNL in Table \ref{tab:summary}). The black lines represent the utilization upper bound, while the green lines show the utilization time series. \\\hspace{\textwidth}
    (c) and (d) are for the maximum amounts of flexibility for power system services with different duration and frequency requirements. (c) Results for the AI-focused HPC data center (Saturn). (d) Results for the general-purpose HPC data center (ORNL). The maximum flexibility amounts are normalized (`Norm. Flex') as ratios of the maximum power of the data center. The heat maps in each column are subject to a specific maximum delay limit (max delay), which is the maximum delay time proportional to the job computing time. ``min'', ``max'', and ``mean'' refer to the minimum, maximum, and mean values across all blocks in each heat map.}
    \label{fig:uti_base_max_amount}
\end{figure}

\subsection{Maximum amount of data center flexibility}
\label{res:size_of_flex}

We first evaluate the maximum amount of flexibility that data centers can provide for a wide range of power system services. Each of these services can be characterized by duration, frequency, and response time \cite{pjm-ancillary-services, storage_joule}. Frequency indicates how often the power system operator activates the service; duration is the length of time that each activation lasts; and response time refers to how quickly a resource delivers the requested flexibility. For example, primary response and tertiary response are two types of power system services that are often used by grid operators to maintain system frequency within an acceptable range following a demand-supply mismatch. Primary response only requires power delivery to be sustained for a few minutes after activation, but may be activated 15,000 times a year \cite{storage_joule}. On the other hand, tertiary response aims to restore the power system to its normal state, and is rarely activated (approximately 20-50 times a year), but when activated may require continuous power delivery for several hours \cite{storage_joule}. It should be noted that although electricity markets will generally have similar power system services, their names, delivery requirements, and detailed procurement mechanisms will vary. This paper focuses only on duration and frequency. Ref. \cite{Zhang2022} showed that data centers can track power adjustment every few seconds, which is generally sufficient for most power system services. 

During the activation periods of each power system service, the data center flexibility is the difference between the baseline power and the power after adjusting the computing workloads. This workload adjustment can be achieved in practice through job preemption \cite{borg, rnuai} and hardware controls \cite{krzywaniak2023dynamic,dutta2018gpu, krzywaniak2022depo} as discussed in the \nameref{sec:method} section. We only consider upward flexibility (i.e., demand reduction), because data centers tend to be highly utilized so downward flexibility potential is very limited. Job completion times could be delayed after adjustment, but we have imposed a maximum delay limit as was done in \cite{anc_datacenter, ding_datacenter_rl, Zhang2022}. In this setting, the maximum flexibility for each power system service is calculated by solving optimization problems for job rescheduling (see \nameref{sec:method}). Due to the computational complexity of solving the optimization problems, our analysis starts by comparing the results for one general-purpose HPC data center (ORNL in Table \ref{tab:summary}) and one AI-focused HPC data center (Saturn in Table \ref{tab:summary}). These two data centers have average utilization rates both at 81\%, and Figs. \ref{fig:uti_base_plot_AI} and \ref{fig:nuti_base_plot_General} shows their baseline utilization time series. Due to the lack of data for data center electric power, we assume a linear relationship between the data center utilization and electric power. This paper considers CPUs as the resource constraint for general-purpose HPC data centers, while GPUs as that for AI-focused HPC data centers due to their importance in AI jobs and higher power consumption \cite{seren}. General-purpose HPC data centers also have some GPU jobs, and their CPU utilization has been considered.
 
Fig. \ref{fig:normalised_flex_AI} shows the maximum amounts of flexibility calculated for power system services with different frequency and duration requirements for the AI-focused HPC data center (Saturn), and \ref{fig:normalised_flex_General} shows the results for the general-purpose HPC data center (ORNL). The time resolution in our optimization problems is set to 15 minutes for computational tractability. Therefore, Figs. \ref{fig:normalised_flex_AI} and \ref{fig:normalised_flex_General} have considered all possible duration and frequency requirements in this setting. These requirements also cover all the 12 typical power system services listed in Ref. \cite{storage_joule}.
Comparing Figs. \ref{fig:normalised_flex_AI} and \ref{fig:normalised_flex_General}, it is apparent that the AI-focused HPC data center provides greater flexibility than the general-purpose HPC data center as the service duration increases above 1 hour. 
This is because the general-purpose HPC data center has more variable baseline utilization (Figs. \ref{fig:uti_base_plot_AI} and \ref{fig:nuti_base_plot_General}), so it is more difficult to sustain demand reduction flexibility for a long time. This difference in utilization patterns is also observed in our analysis for all 14 data centers, as will be illustrated in Fig. \ref{fig:all_HPC_corr}.
Note that, Figs. \ref{fig:normalised_flex_AI} and \ref{fig:normalised_flex_General} show that the general-purpose HPC data center has slightly greater flexibility for short-duration services. In fact, this is because our job partition strategy (see \nameref{sec:method}) underestimates the flexibility of the AI-focused HPC data center with more long-computing-time jobs. As will be illustrated in Fig. \ref{fig:size_vs_numdays}, these two data centers have similar amounts of flexibility for short-duration services with a sufficiently long optimization horizon.

\begin{figure*}
    \centering
    \subfloat[]{%
    \includegraphics[width=0.5\linewidth]{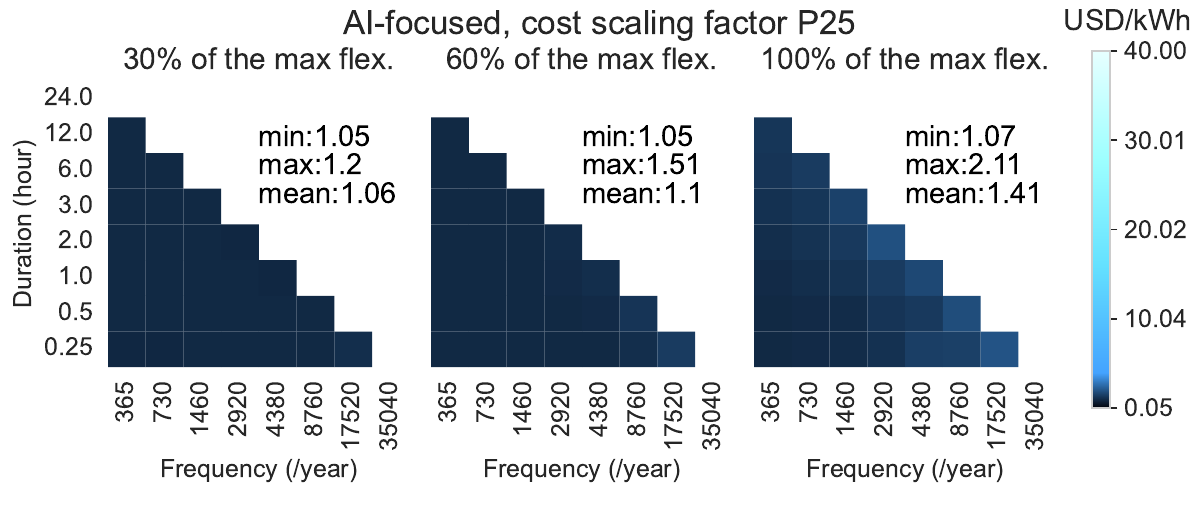}
      \label{fig:unit_cost_AI_P25}
    }
    \subfloat[]{%
      \includegraphics[width=0.5\linewidth]{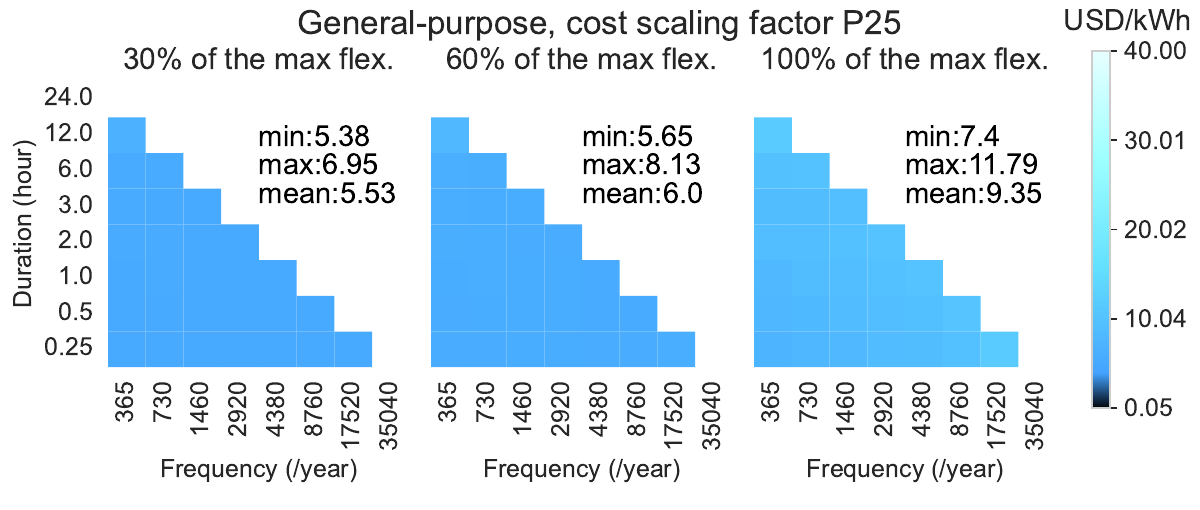} \label{fig:unit_cost_General_P25}
    }\\
    \subfloat[]{%
    \includegraphics[width=0.5\linewidth]{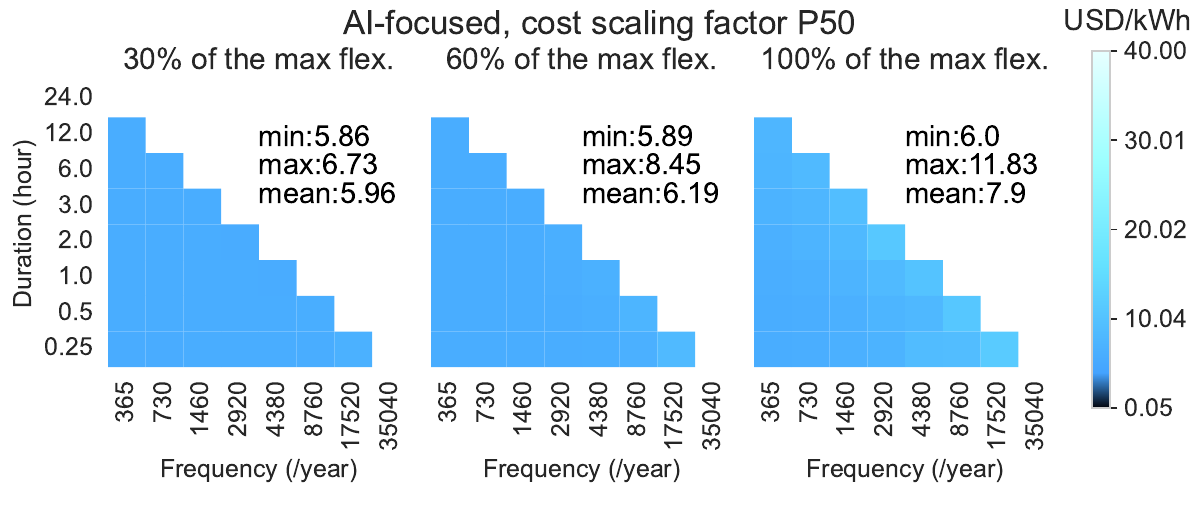}
      \label{fig:unit_cost_AI_P50}
    }
    \subfloat[]{%
      \includegraphics[width=0.5\linewidth]{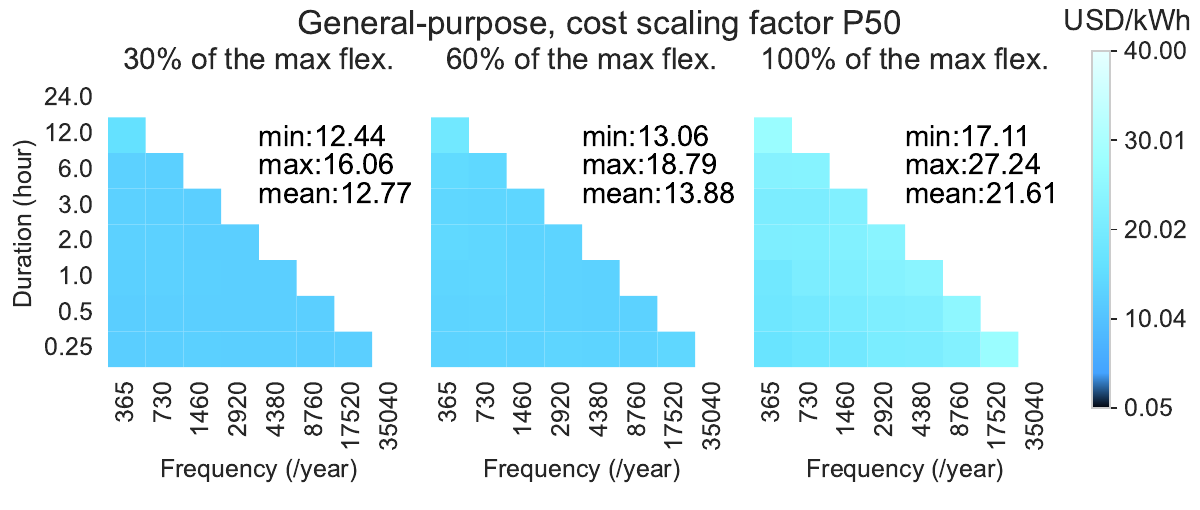} \label{fig:unit_cost_General_P50}
    }\\
    \subfloat[]{%
    \includegraphics[width=0.5\linewidth]{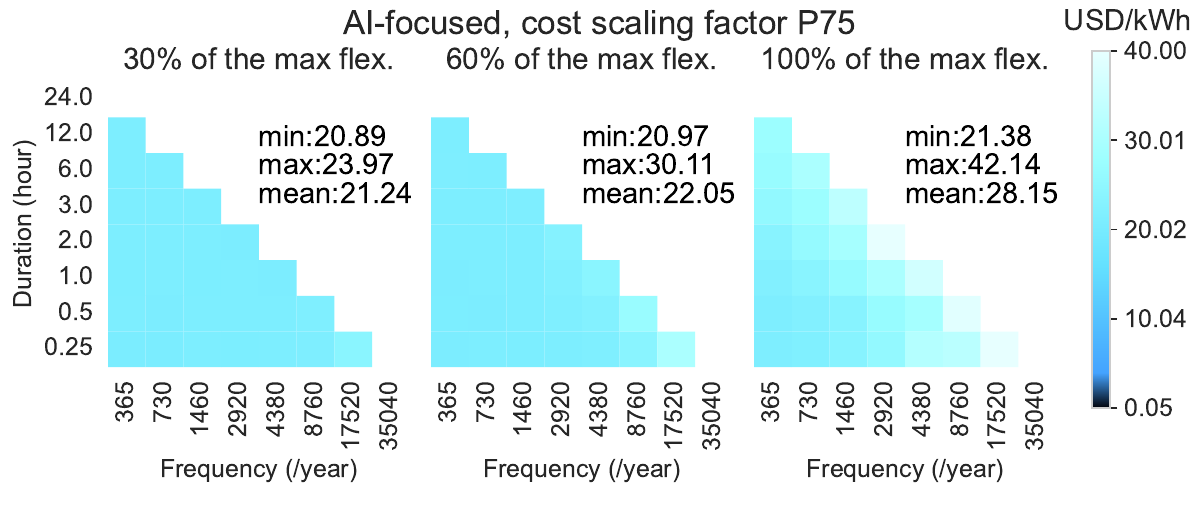}
      \label{fig:unit_cost_AI_P75}
    }
    \subfloat[]{%
      \includegraphics[width=0.5\linewidth]{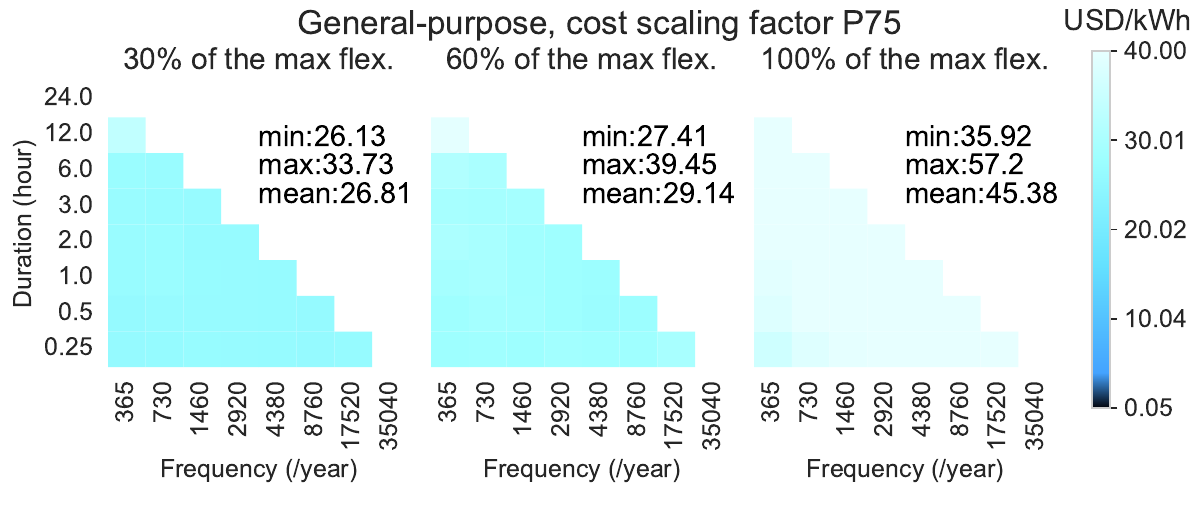} \label{fig:unit_cost_General_P75}
    }\\
    \subfloat[]{%
      \includegraphics[width=1\linewidth]{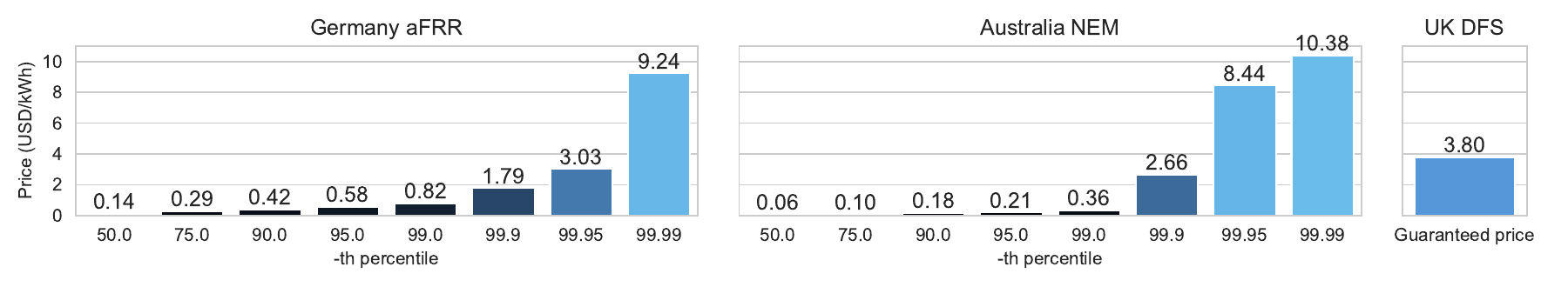} \label{fig:market_prices}
    }
    
  \vspace{-2 mm}
  \caption{Average flexibility cost and power system service prices. In (a)-(f), each heatmap displays the average cost of providing different percentages of the maximum amounts of flexibility evaluated in Figs. \ref{fig:normalised_flex_AI} and \ref{fig:normalised_flex_General}. The maximum delay limit is set to 20\%. ``min'', ``max'', and ``mean'' refer to the minimum, maximum, and mean values across all blocks in each heatmap. Plots (a), (c), and (e) correspond to the AI-focused HPC data center (Saturn) under 25-th (P25), 50-th (P50), and 75-th (P75) percentiles of the cost scaling factor, estimated using data from Google Cloud, AWS, and Oracle. Plots (b), (d), and (f) correspond to the general-purpose HPC data center (ORNL). Plot (g) shows percentiles of real-world power system service prices from Germany's aFRR \cite{aFRR_German}, Australia's NEM \cite{AEMO_price_data}, and UK's DFS \cite{dfs}.}
  \label{fig:unit_cost}
  \vspace{-2 mm}
\end{figure*}

\begin{figure*}[tb]
    \centering
    \includegraphics[width=1\linewidth]{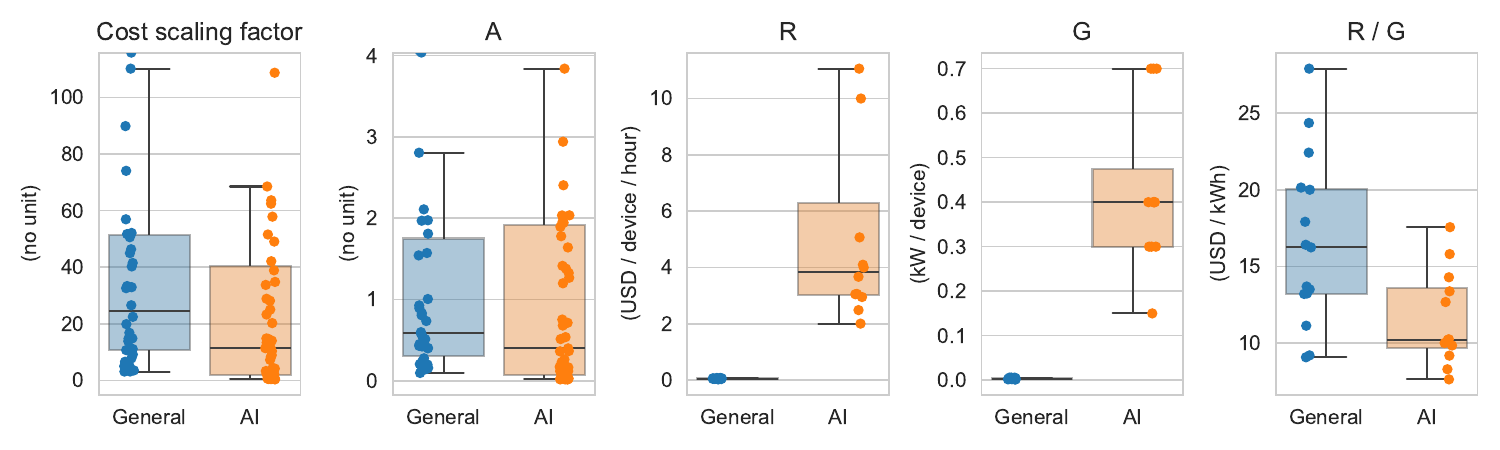}
    \vspace{-9 mm}
    \caption{Samples of the cost scaling factor and the elements ($A$, $R$, and $G$) involved in its computation (see Eq. \eqref{eq:scale_ACoF_nodq}). These samples are derived based on data of computing rental options on Google Cloud, AWS, and Oracle. The 25th, 50th, and 75th percentiles of the cost scaling factor for general-purpose HPC data centers (General) are 10.60, 24,49, and 51.42 respectively. The 25th, 50th, and 75th percentiles of the cost scaling factor for AI-focused HPC data centers (AI) are 2.03, 11.37, and 40.48 respectively. 
    \\\hspace{\textwidth} $A$ is the price reduction coefficient that represents the proportionate price reduction in response to a certain proportion of job delay. $R$ is the hourly price of a single virtual CPU (vCPU, for general-purpose HPCs) or a single GPU (for AI-focused HPCs). $G$ is the power of a single vCPU or a single GPU. These parameters are interpreted in the \nameref{sec:method} section. On most cloud platforms, CPUs are rented on the basis of vCPUs, which typically represents one thread of a physical CPU. We follow their convention here. The power of a vCPU ($G$) is calculated by dividing the rated power of the physical CPU by the number of vCPUs. Note that, in the plots of $R$, $G$, and $R/G$, when a set of computing rental options are essentially renting different portions of the same type of machine, we only keep one option to avoid over-counting the same machine for its power and price parameter.}
    \label{fig:interpret_cost_scaling_factor}
    \vspace{-2 mm}
\end{figure*}

\begin{figure}
    \centering
    \subfloat[]{%
    \includegraphics[width=\linewidth]{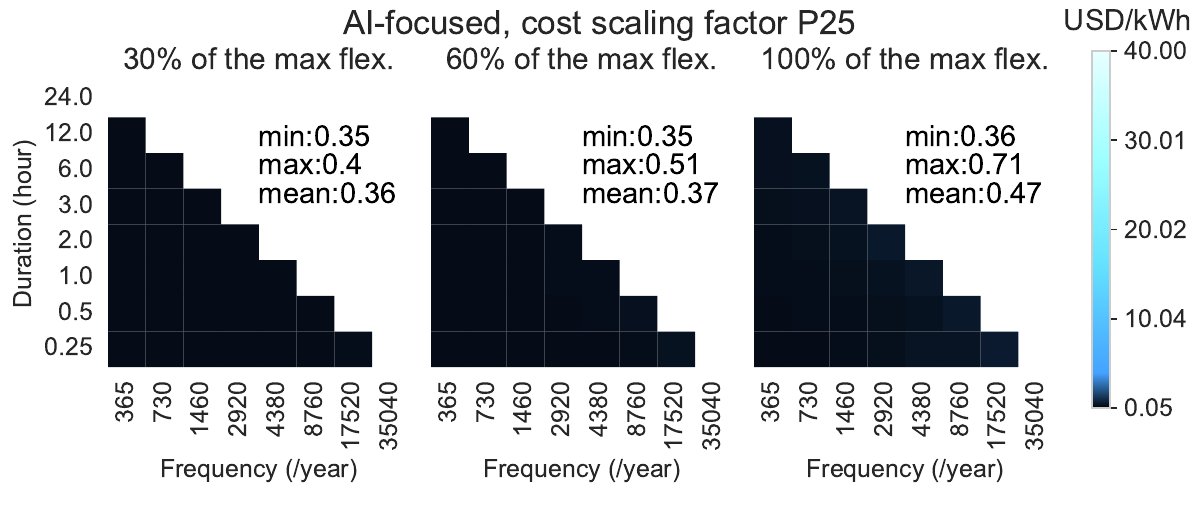}
      \label{fig:unit_cost_AI_lambda_P25}
    }\\
    \subfloat[]{%
    \includegraphics[width=\linewidth]{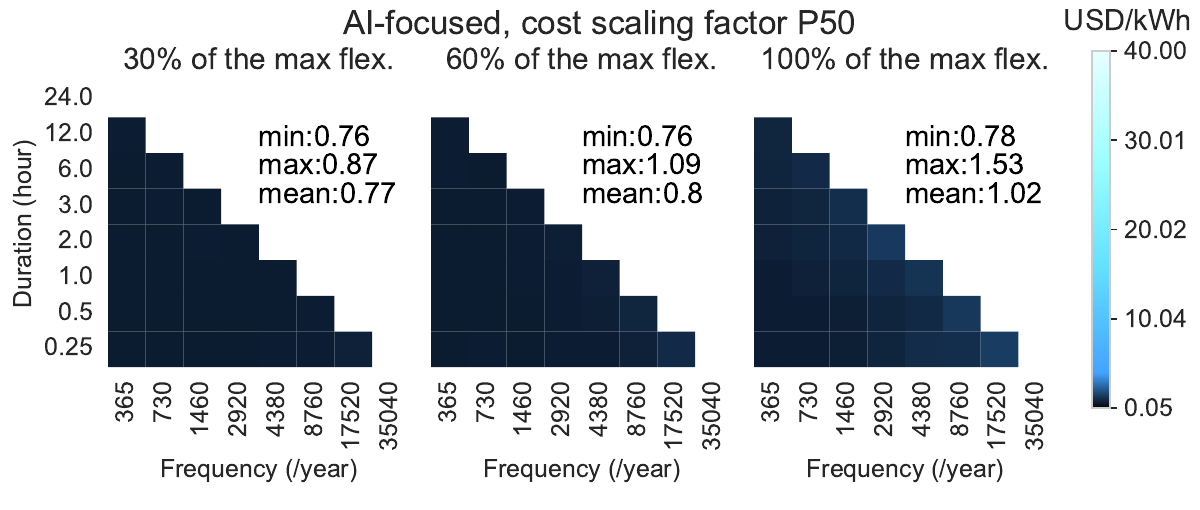}
      \label{fig:unit_cost_AI_lambda_P50}
    }\\
    \subfloat[]{%
    \includegraphics[width=\linewidth]{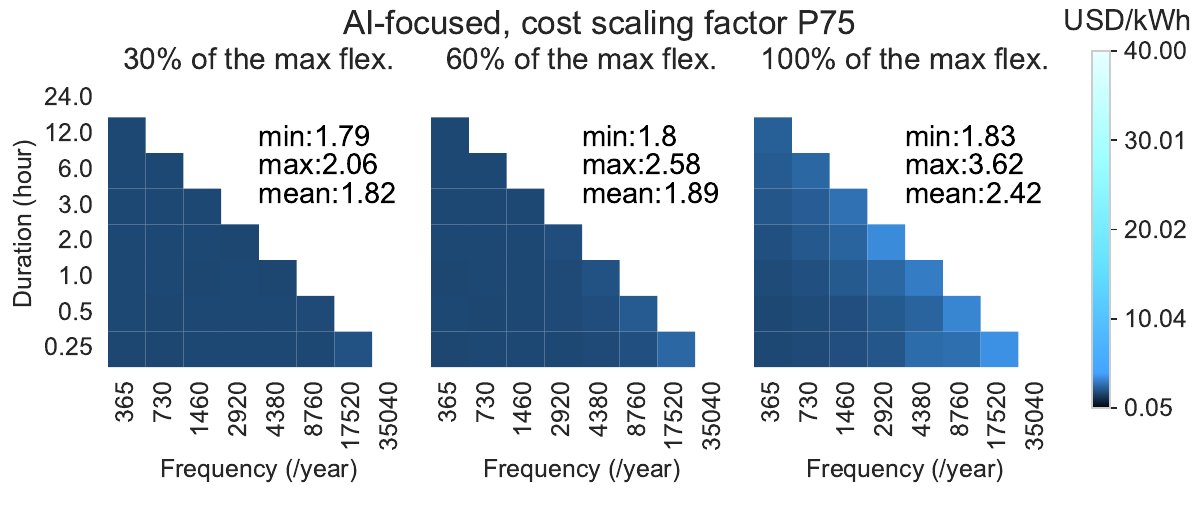}
      \label{fig:unit_cost_AI_lambda_P75}
    }
    
  \vspace{-2 mm}
  \caption{Average flexibility cost under the cost scaling factor derived by Lambda GPU Cloud data \cite{lambda_gpu_cloud}. These cost results are for the AI-focused HPC data center (Saturn) when providing different percentages of the maximum amounts of flexibility evaluated in Fig. \ref{fig:normalised_flex_AI}. The maximum delay limit is set to 20\%. ``min'', ``max'', and ``mean'' refer to the minimum, maximum, and mean values across all blocks in each heatmap. (a), (b), and (c) are for 25-th, 50-th, and 75-th percentiles of the cost scaling factor respectively.}
  \label{fig:unit_cost_lambda}
  \vspace{-2 mm}
\end{figure}

\subsection{Cost of data center flexibility}
\label{res:cost_flex}

We assume that data centers charge a price for each computing job, as is the case with cloud computing platforms. We further assume that the price will be lower if there is greater delay in job completion, due to reduced timeliness. Thus, the total cost for the data center providing a particular power system service is the sum of the price reduction incurred by the delays across all jobs given optimal job scheduling. We propose a linear cost model such that the price reduction for each job is equal to the product of the original price, the job delay proportion, and a price reduction coefficient. This model is formally defined as Eq. \eqref{eq:comp_cost_def} in the \nameref{sec:method} section. Note that the proposed cost model is not limited to cloud computing data centers, because other data centers may also implement pricing schemes for internal usage; delays in computing-related projects (e.g., developing systems such as ChatGPT) could similarly be monetized with appropriate financial data and analysis---however, these data are typically only available for cloud computing data centers. Nevertheless, we anticipate that any deviations from our current cost estimates would be minor for the same computing hardware, as such hardware (e.g., GPUs or CPUs) would otherwise not be rentable through cloud computing data centers.

We focus on \emph{average flexibility cost}, which is the total flexibility cost divided by the total shifted energy during the service activation periods in kWh (the product of the flexibility amount, duration, and frequency). If the average cost is less than the price of the power system service, then providing that service is profitable. In our proposed cost model, the average flexibility cost is calculated using optimal job scheduling. Therefore, the cost is calculated by solving an optimization problem which is similar to how we calculated the maximum amount of flexibility. The objective is to minimize the flexibility cost, with constraints specifying the amount of flexibility provision. This optimization problem is solved for a nominal set of data center parameters, including the power of the computing devices, the computing price, and the price reduction coefficient. Thanks to the use of the linear cost model, we can quickly calculate the average flexibility cost under other parameter settings by multiplying the corresponding cost scaling factors (see \nameref{sec:method}). The cost scaling factor represents the ratio of the average flexibility cost under a specific data center parameter setting to that under the nominal setting. Furthermore, the cost scaling factor that reflects real-world data centers can be estimated by comparing the price of one machine with another slower and cheaper machine using real-world data, as detailed in \nameref{sec:method}. Based on the above, we collect the price, power, and computing speed information of all HPC CPU and GPU computing rental options from three large-scale cloud computing platforms: Google Cloud, Amazon AWS, and Oracle. These data are provided in our supplementary material. We collected 38 samples of the cost scaling factor for general-purpose HPC data centers, and 55 samples for AI-focused HPC data centers. These samples provide an indication of the real-world cost characteristics of data centers.

Figs. \ref{fig:unit_cost_AI_P25}--\ref{fig:unit_cost_General_P75} show the average flexibility cost for various power system services when data centers provide different percentages of their maximum flexibility (as evaluated in Figs. \ref{fig:normalised_flex_AI} and \ref{fig:normalised_flex_General}). These figures also display results under different percentiles of the cost scaling factor derived from the samples. It can be seen that the AI-focused HPC data center has lower flexibility cost than the general-purpose HPC data center for all power system services and all percentiles of the cost scaling factor. Specifically, when the cost scaling factor is at the median (P50), the AI-focused HPC data center shows a lower flexibility cost of at least 50\%. One reason for the lower flexibility cost is the smaller cost scaling factor of AI-focused HPC data centers. Fig. \ref{fig:interpret_cost_scaling_factor} shows the distribution of the cost scaling factor and its components. One interesting observation is that, although GPUs (representing AI-focused HPC data centers) are more expensive than CPUs (representing general-purpose HPC data centers), the GPU power consumption is also large, which leads to a lower price-to-power ratio than CPUs, which in turn brings a lower cost scaling factor to AI-focused HPC data centers.

To evaluate the financial profitability of providing power system services, we collect prices of power system services in the UK, Australia, and Germany. In the winters of 2022/23 and 2023/24, the UK's Demand Flexibility Service (DFS) offered a guaranteed acceptance price at 3 GBP/kWh (around 3.8 USD/kWh) for flexibility during stressful system periods \cite{dfs}. 34 test events and 4 live events were performed. For Australia, we collect price data from the National Energy Market (NEM) for 2022-2024 \cite{AEMO_price_data}. For Germany, we collect 2022-2024 prices of automatic frequency restoration reserve (aFRR), a secondary reserve power system service common among Europe \cite{aFRR_German}. Fig. \ref{fig:market_prices} shows the percentiles of these prices. Comparing the price and cost results, it can be seen that the AI-focused HPC data center can achieve profitability under the following conditions: 1) when the cost scaling factor is at the 25th percentile and prices are at or above the 99.9th percentiles for aFRR, NEM, and UK DFS; or 2) when the cost scaling factor is at the 50th percentile and prices are at the 99.99th percentile for aFRR and at or above the 99.95th percentile for NEM. Although rare, these price events underscore the high value of certain power system services, which could be profitable for data centers to provide.

When gathering data from various cloud providers, we found that Lambda GPU Cloud \cite{lambda_gpu_cloud}, which specializes in GPU rentals, offers GPU options approximately 70\% cheaper than those from Google Cloud, AWS, and Oracle. The flexibility cost associated with the AI-focused HPC data center is consequently reduced when using a cost scaling factor based on Lambda GPU Cloud data, as shown in Fig. \ref{fig:unit_cost_lambda}. By comparing Fig. \ref{fig:unit_cost_lambda} and Fig. \ref{fig:market_prices}, it can be seen that the profitability of the AI-focused HPC data center can be achieved even under the 75th percentile of the cost scaling factor. This finding indicates that the profitability varies between data centers with different perceived value in computing. Data centers with lower perceived value in computing might achieve higher profits in providing power system services.

\begin{figure*}
  \centering

  \subfloat[ ]{\includegraphics[width=.5\linewidth]{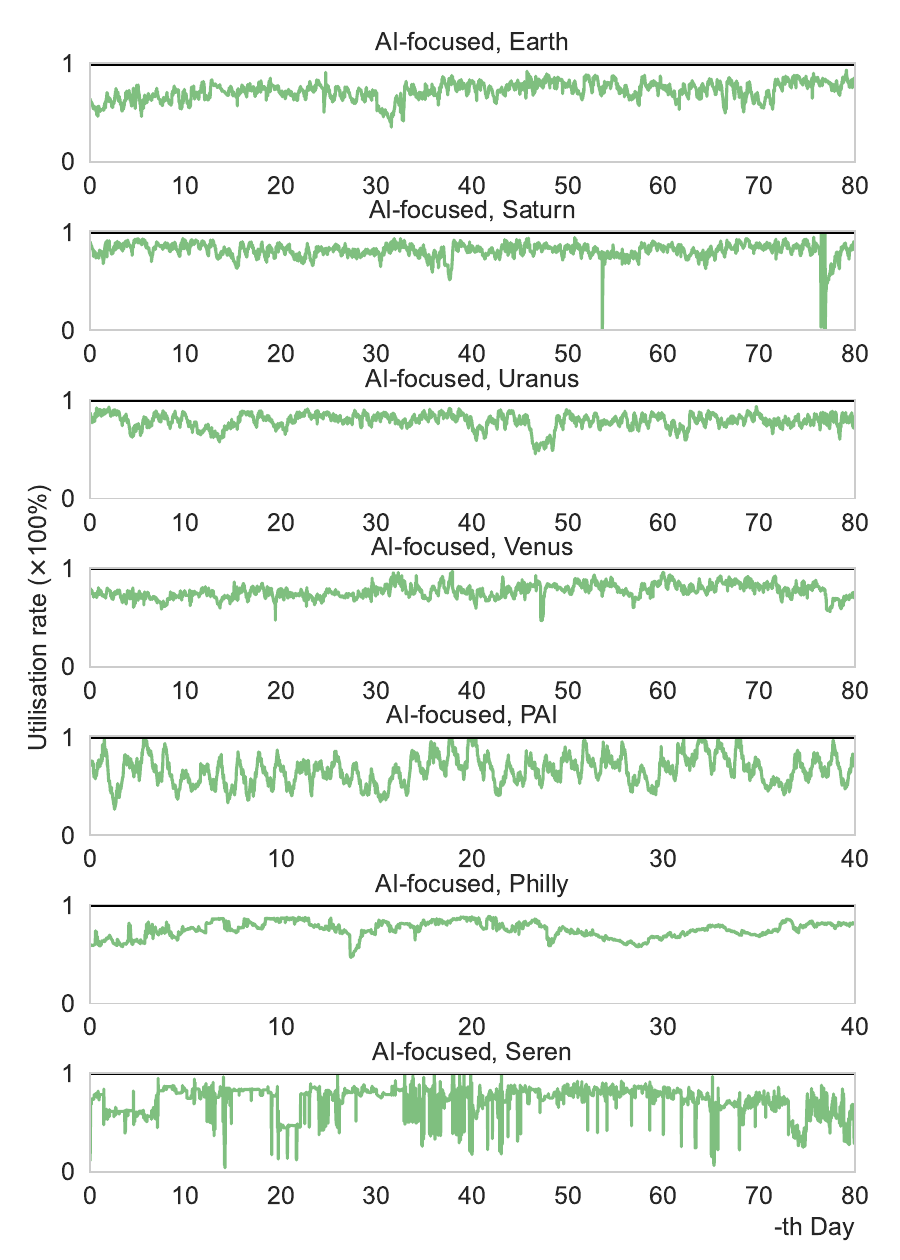}\label{fig:uti_base_AI}}\hfill
  \subfloat[ ]{\includegraphics[width=.5\linewidth]{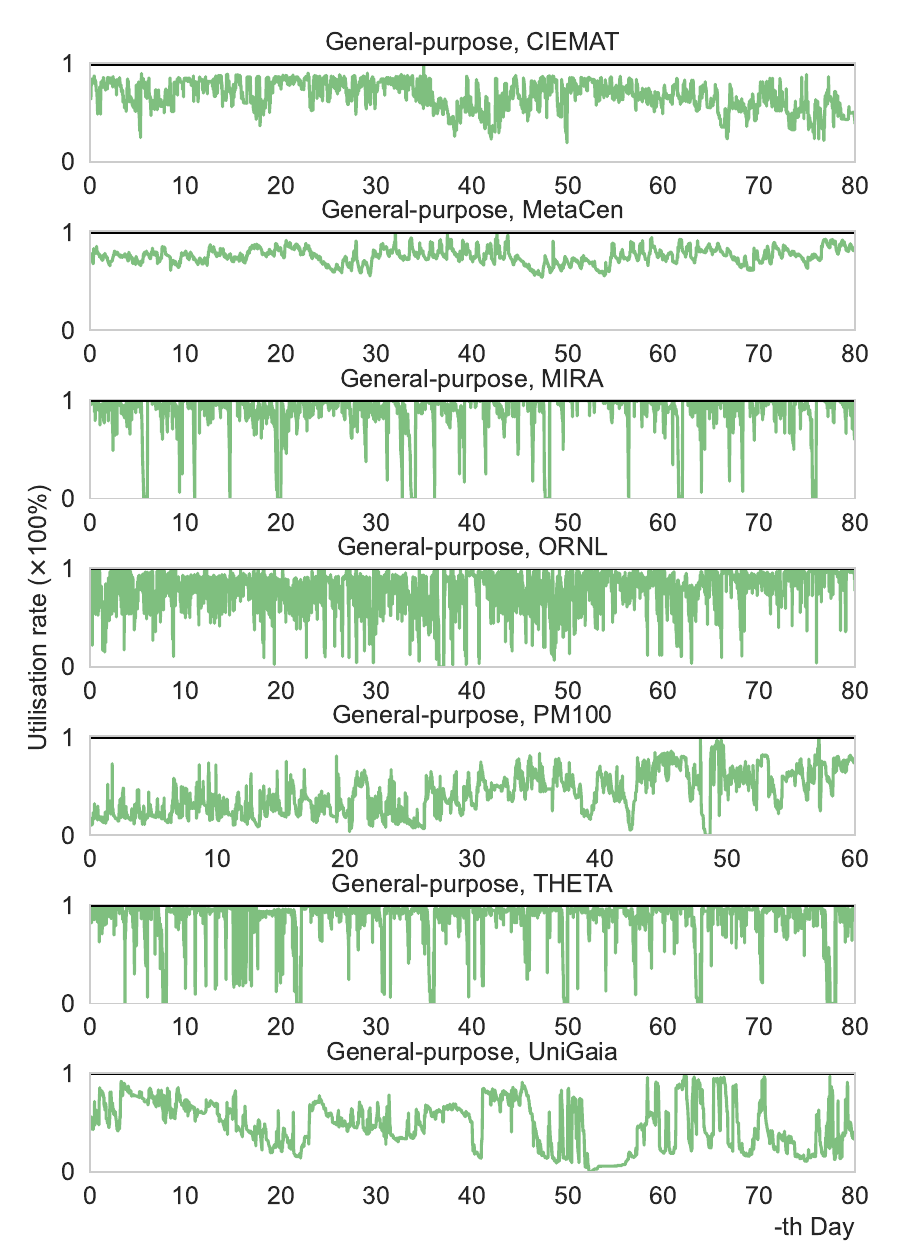}\label{fig:uti_base_General}}

  \subfloat[ ]{\includegraphics[width=.5\linewidth]{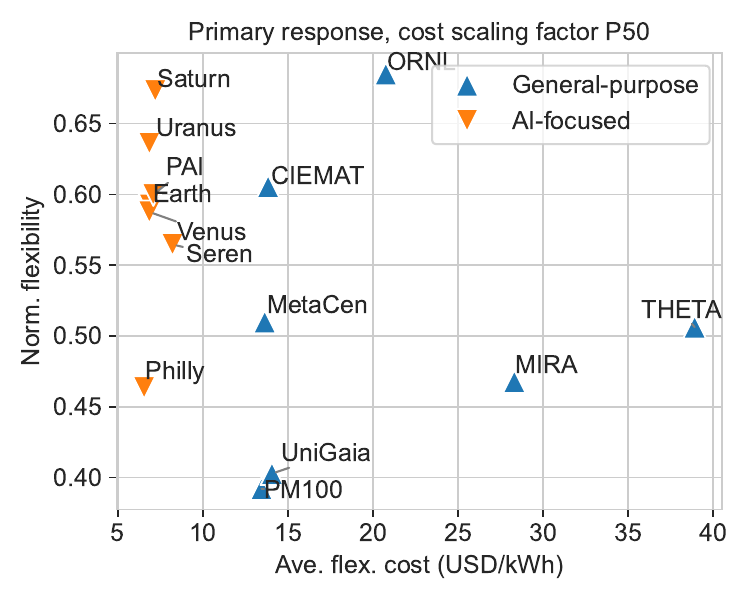}\label{fig:flex_vs_cost_dur0.25freq80_P50}}\hfill
  \subfloat[ ]{\includegraphics[width=.5\linewidth]{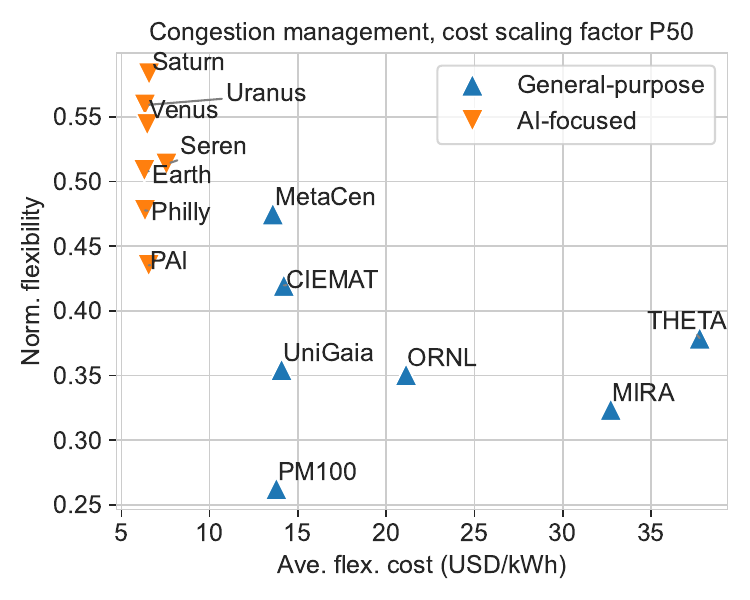}\label{fig:flex_vs_cost_dur2.0freq10_P50}}
   \vspace{ -6mm}
  \caption{Analysis of all the 14 HPC data centers. \\\hspace{\textwidth} (a) Baseline utilization time series (green lines) of the 7 AI-focused HPC data centers. (b) Baseline utilization time series (green lines) of the 7 general-purpose HPC data centers. The black lines refer to the utilization upper bound.  \\\hspace{\textwidth} (c) and (d): The normalized maximum amount of flexibility (Norm. flexibility) versus the average flexibility cost (ave. flex. cost) when providing 100\% of the maximum flexibility for two power system services. We use the 50-th percentile (P50) of the cost scaling factor (estimated using data from Google Cloud, AWS, and Oracle). (c) Results for providing primary response (duration of 0.25 hours and frequency of 2920 times/year); (d) Results for providing congestion management (duration of 2 hours and frequency of 365 times/year). Dots closer to the top-left corner indicate better flexibility providers in terms of greater flexibility and lower cost.}
  \label{fig:all_HPC_analysis}
  \vspace{-2 mm}
\end{figure*}


\begin{figure}
  \centering
  \subfloat[ ]{\includegraphics[width=1\linewidth]{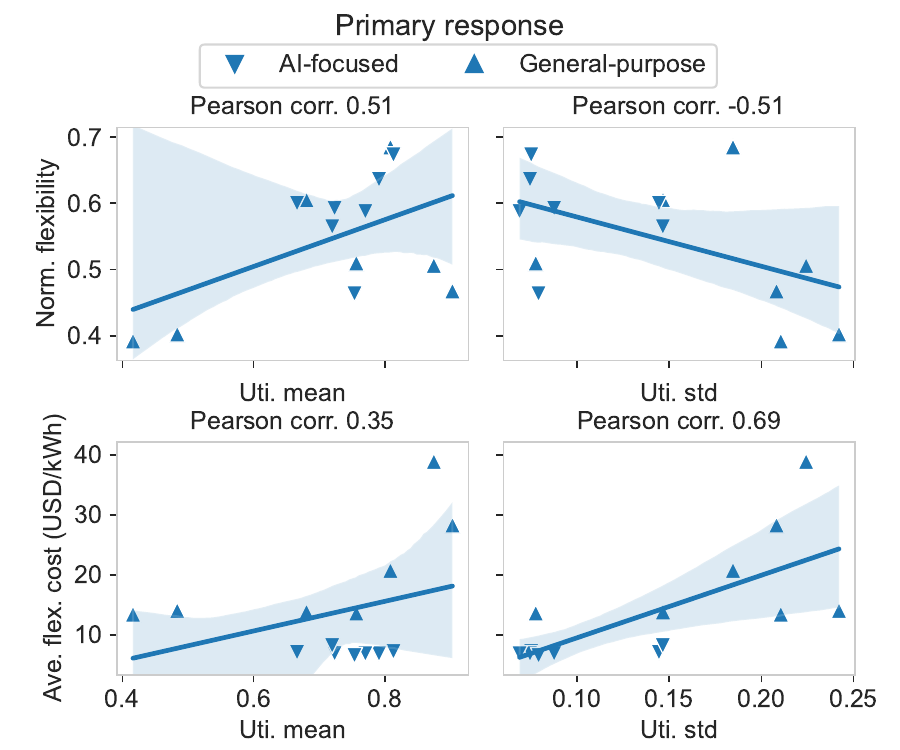}\label{fig:corrplot_dur0.25freq80}}\hfill
  \subfloat[ ]{\includegraphics[width=1\linewidth]{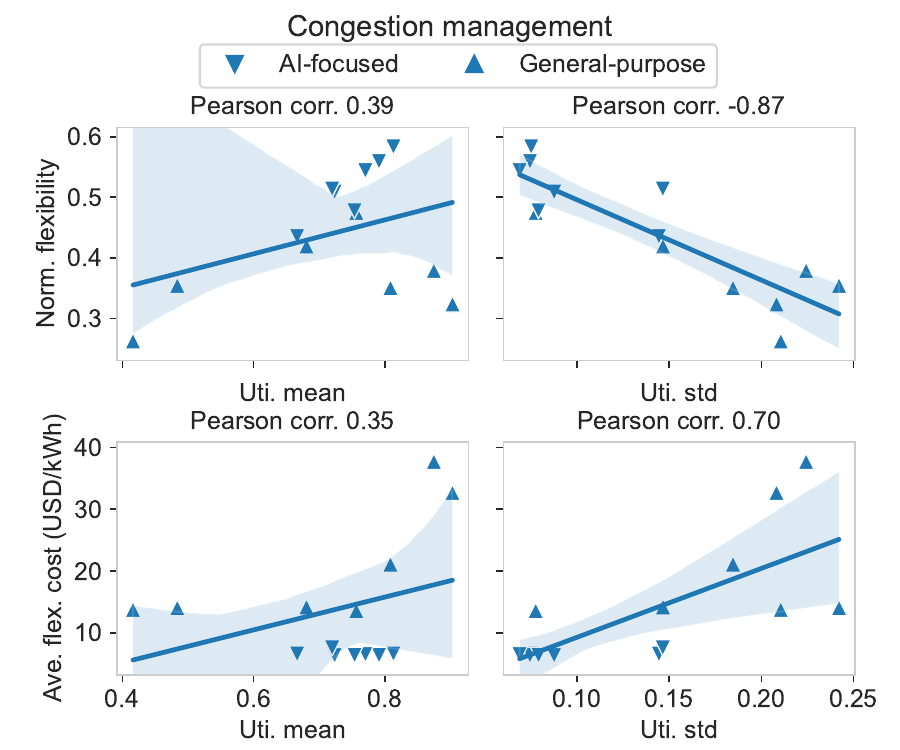}\label{fig:corrplot_dur2.0freq10}}\hfill

\subfloat[ ]{\includegraphics[width=.8\linewidth]{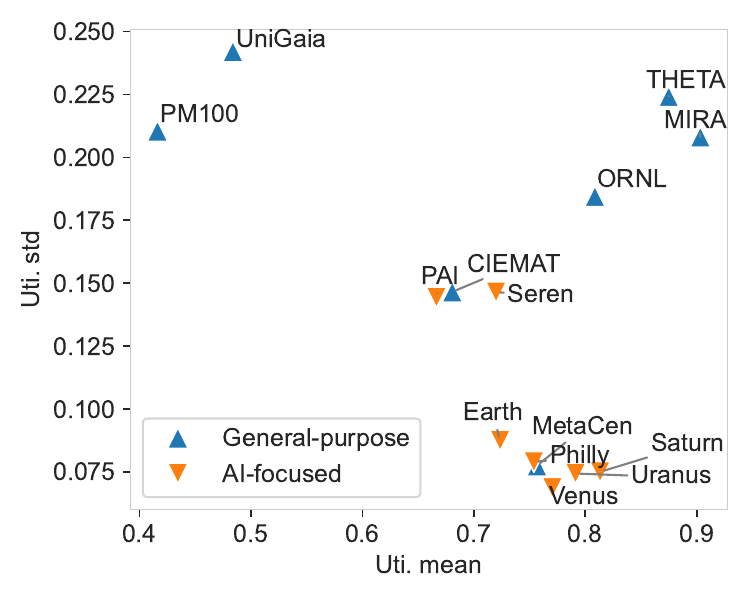}\label{fig:mean_vs_std}}
  
   \vspace{-2 mm}
  \caption{Correlation analysis of the flexibility results and data center utilization patterns. The cost scaling factor is set to the 50th percentile (estimated using data from Google Cloud, AWS, and Oracle). (a) Correlation between the normalized maximum amount of flexibility (Norm. flexibility) for primary response, the average flexibility cost (Ave. flex. cost) for primary response, the mean utilization rate (Uti. mean), and the standard deviation of utilization (Uti. std). Regression lines with confidence intervals (shaded areas) are plotted for highlighting the trends. (b) Correlation results where the flexibility and cost are for congestion management. (c) The mean utilization rate and the standard deviation of utilization for all the 14 data centers.}
  \label{fig:all_HPC_corr}
\end{figure}

\subsection{Generalizability of the finding}
\label{sec:reasons_and_gel}

To generalize the previous findings, this section analyzes the 7 AI-focused HPC data centers and 7 general-purpose HPC data centers in Table \ref{tab:summary} for two typical power system services: One is primary response characterized by short duration and high frequency (0.25 hours and 2920 times a year), while the other is congestion management characterized by long duration and low frequency (2 hours and 365 times a year). It should be noted that these two power system services may have different duration and frequency requirements within a certain range. The specific ranges for these requirements are detailed in \cite{storage_joule}.

Figs. \ref{fig:uti_base_AI} and \ref{fig:uti_base_General} display the utilization time series of the 14 HPC data centers. We select 80 consecutive days of job records from the most recent year to ensure up-to-dateness, except a few with only 40 or 60 days after data cleaning. The details of the data processing are given in \nameref{sec:method}.
Fig. \ref{fig:flex_vs_cost_dur0.25freq80_P50} shows the maximum amount of flexibility versus the average flexibility cost when providing primary response, while Fig. \ref{fig:flex_vs_cost_dur2.0freq10_P50} shows that for providing congestion management. When providing primary response, it can be seen that AI-focused HPC data centers have lower flexibility cost than all general-purpose HPC data centers. However, it should also be noted that there are two general-purpose HPC data center (ORNL and CIEMAT) that have maximum flexibility approximately equal to or greater than some AI-focused HPC data centers. When providing congestion management, AI-focused HPC data centers are more advantageous: All AI-focused HPC data centers have at least 50\% lower flexibility cost than all general-purpose HPC data centers, and only one AI-focused HPC data center (PAI) shows approximately 7\% less flexibility than a specific general-purpose HPC data center (MetaCen).

It should be noted that Figs. \ref{fig:flex_vs_cost_dur0.25freq80_P50} and \ref{fig:flex_vs_cost_dur2.0freq10_P50} are based on the 50th percentile of the cost scaling factor. Our supplementary file provides plots under the 25th and 75th percentiles, and the conclusion of this section remains unchanged.


\subsection{Impact of job patterns on the maximum flexibility and flexibility cost}

In addition to the different cost scaling factors illustrated in Fig. \ref{fig:interpret_cost_scaling_factor}, the different utilization patterns also contribute to the greater flexibility and lower cost of AI-focused data centers compared to general-purpose data centers. Fig. \ref{fig:corrplot_dur0.25freq80} illustrates the correlation between the normalized maximum amount of flexibility for the primary response service, the average flexibility cost for the primary response service, the mean baseline utilization rate and the standard deviation of baseline utilization; Fig. \ref{fig:corrplot_dur2.0freq10} shows that for the congestion management service. The first column of Figs. \ref{fig:corrplot_dur0.25freq80} and \ref{fig:corrplot_dur2.0freq10} shows a positive correlation between the mean baseline utilization and the normalized maximum amount of flexibility, and between the mean baseline utilization and the average flexibility cost. This is intuitive as a higher mean utilization rate implies a higher baseline power, leading to greater flexibility. At the same time, a highly utilized data center faces more difficulty in restoring jobs disrupted by flexibility provision, resulting in higher cost.

The second column of Figs. \ref{fig:corrplot_dur0.25freq80} and \ref{fig:corrplot_dur2.0freq10} shows a negative correlation between the standard deviation of utilization and the normalized maximum amount of flexibility, which is strong (-0.87) for the congestion management service. This is because data centers with variable utilization time series (indicating variable power demand) are less capable of sustaining flexibility for a long time. We also find a strong positive correlation between the standard deviation of utilization and the average flexibility cost. This is because variant utilization time series can have more frequent utilization spikes, which impede timely job restoration and thus increase flexibility cost.

Finally, Fig. \ref{fig:mean_vs_std} shows the mean utilization and the standard deviation of utilization for all data centers. AI-focused HPC data centers tend to have relatively high utilization rates but low variance, enabling them to provide greater flexibility at lower cost especially for long-duration services such as congestion management, as has been illustrated in Fig. \ref{fig:flex_vs_cost_dur2.0freq10_P50}.

Figs. \ref{fig:corrplot_dur0.25freq80} and \ref{fig:corrplot_dur2.0freq10} use the 50th percentile of the cost scaling factor. Our supplementary file provides figures under the 25th and 75th percentiles. The same conclusion can still be derived.

\subsection{Dynamic quota for greater flexibility and lower cost}

\begin{figure}[tb]
  \centering

  \subfloat[ ]{\includegraphics[width=1\linewidth]{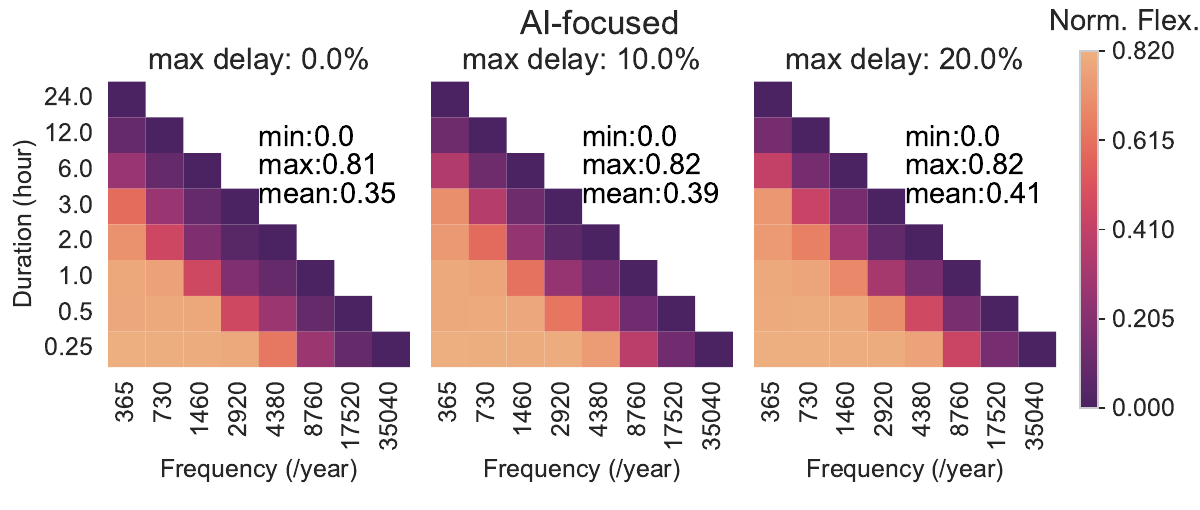}\label{fig:normalised_flex_AI_dq0.5}}\hfill
    
  \subfloat[ ]{\includegraphics[width=1\linewidth]{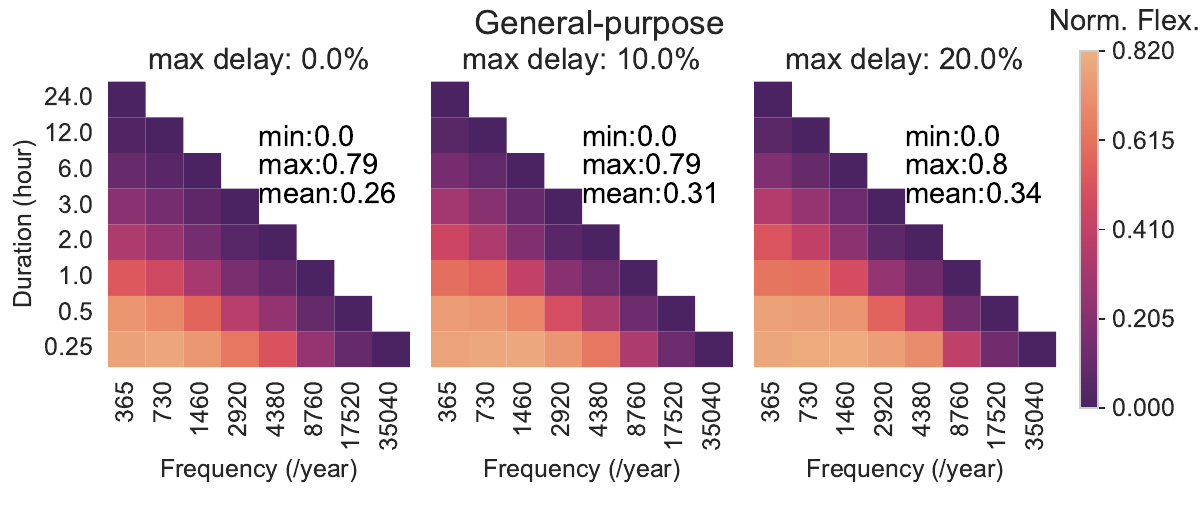}\label{fig:normalised_flex_General_dq0.5}}

  \subfloat[ ]{\includegraphics[width=1\linewidth]{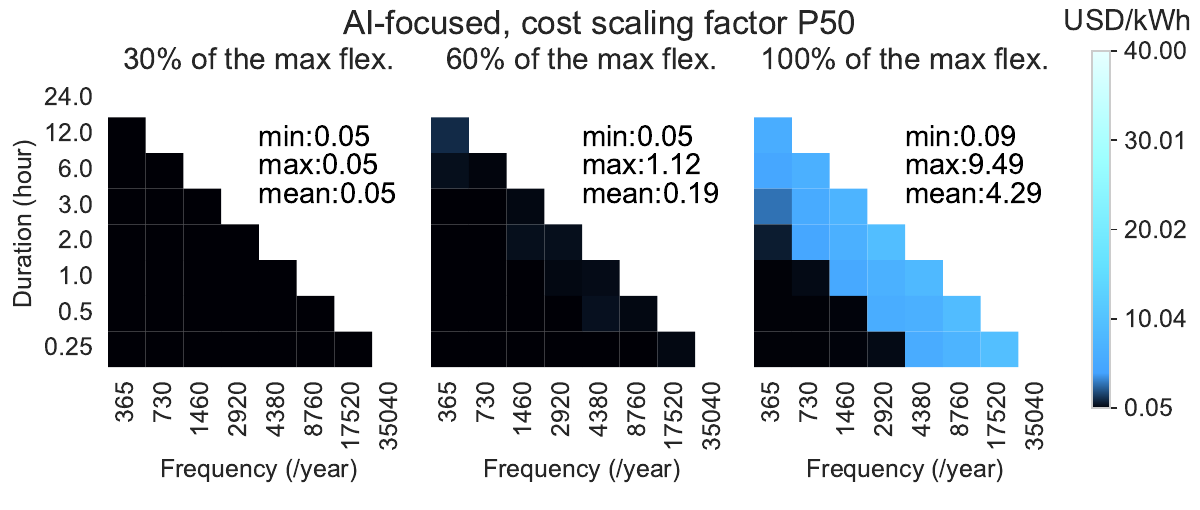}\label{fig:unit_cost_AI_dq0.5}}\hfill
  
  \subfloat[ ]{\includegraphics[width=1\linewidth]{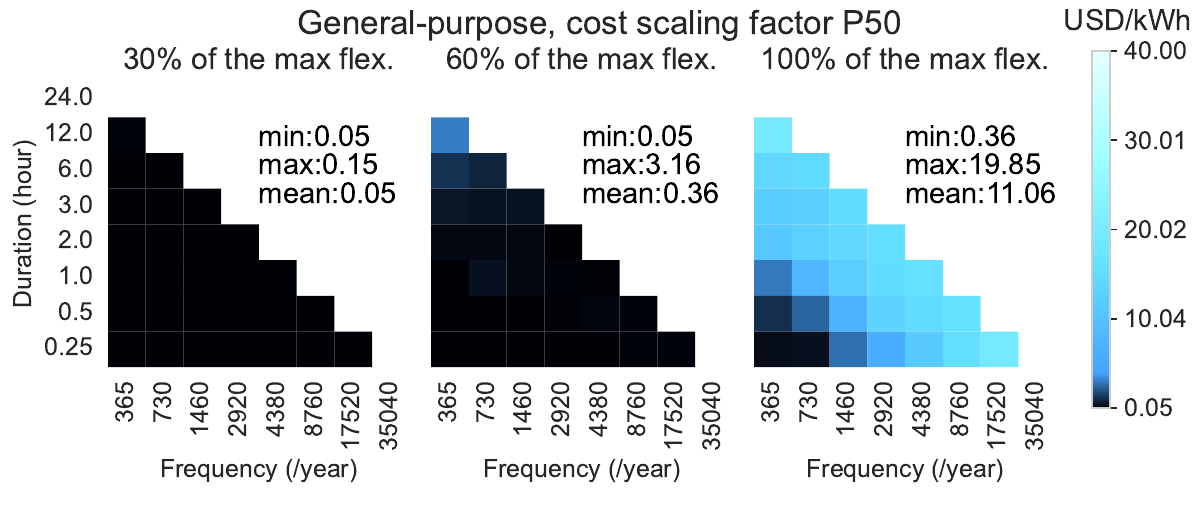}\label{fig:unit_cost_General_dq0.5}}\hfill
   \vspace{ -6mm}
  \caption{The maximum amount of flexibility and the flexibility cost with the dynamic quota opportunity.
  \\\hspace{\textwidth} (a) The maximum amounts of flexibility for the AI-focused HPC data center (Saturn). (b) The maximum amounts of flexibility for the general-purpose HPC data center (ORNL). \\\hspace{\textwidth}
  (c) and (d): The average cost of providing different percentages of the maximum amounts of flexibility evaluated in (a) and (b). (c) The average flexibility cost for the AI-focused HPC data center (Saturn). (d) The average flexibility cost for the general-purpose HPC data center (ORNL). We use the 50th percentile of the cost scaling factor (estimated using data from Google Cloud, AWS, and Oracle).
  \\\hspace{\textwidth}``min'', ``max'', and ``mean'' refer to the minimum, maximum, and mean values across all blocks in each subplot. Using 100\% extra computing resources is assumed to have 50\% computing speed-up.}
  \label{fig:flex_cost_dq}
  \vspace{-2 mm}
\end{figure}

Certain computing jobs, such as AI training with batch computation, are parallelizable. Therefore, as long as data center computing resources are not always 100\% utilized, they can leverage unused resources to speed up computations, reducing job delays and thus flexibility cost. This is called ``dynamic quota'' as in \cite{run_ai_dq_name} and is used by data center schedulers \cite{rnuai}. Figs. \ref{fig:normalised_flex_AI_dq0.5} and \ref{fig:normalised_flex_General_dq0.5} show the maximum amounts of flexibility for the two data centers in Figs. \ref{fig:normalised_flex_AI} and \ref{fig:normalised_flex_General}, where 100\% extra computing resources can only increase 50\% of computing speed. It can be seen that the maximum flexibility increases noticeably compared to Figs. \ref{fig:normalised_flex_AI} and \ref{fig:normalised_flex_General}, and there is now significant flexibility even when no computing delay is allowed (this is in contrast to the results without dynamic quota in Figs. \ref{fig:normalised_flex_AI} and \ref{fig:normalised_flex_General} where there is no flexibility when no delay is allowed). 

Figs. \ref{fig:unit_cost_AI_dq0.5} and \ref{fig:unit_cost_General_dq0.5} show the flexibility cost with the dynamic quota opportunity. Using extra resources for computational speed-up imposes additional flexibility cost due to higher energy usage, which are calculated based on a wholesale energy market price of 0.05 USD/kWh. Figs. \ref{fig:unit_cost_AI_dq0.5} and \ref{fig:unit_cost_General_dq0.5} show that the average flexibility cost can be as low as 0.05 USD/kWh, which is close to grid energy storage \cite{storage_joule}. This is because data centers with dynamic quota can provide power system services without incurring job delays.

\section{Discussion}
\label{sec:discussion}

This paper evaluated the maximum flexibility that data centers can provide through job scheduling, and proposed a method to estimate the associated flexibility cost. Based on real-world datasets of 14 data centers, we found that AI-focused HPC data centers can provide greater flexibility at lower cost, especially for power system services with longer duration requirements such as congestion management. By comparing the flexibility cost and the real-world power system service prices, we illustrated the financial profitability of data centers providing flexibility. Finally, where the dynamic quota feature is applicable, it can further increase the flexibility of data centers and reduce the associated flexibility cost.

Our findings have implications for several stakeholders. For power system operators who want to manage increasing electricity demand while deferring expensive grid infrastructure upgrades, they can design more targeted collaboration strategies with data centers based on their differing capabilities for providing specific power system services. For example, AI-focused HPC data centers can be prioritized when securing power system services that require long duration. Our findings also have implications for data center operators, who may believe that they should not disrupt ``high-value'' computing jobs to provide ``low-value'' power system services. Our findings show that providing power system services could bring extra net profit to data centers. It is worth noting that the value of flexibility can persist even when data centers are constructed in areas with sufficient connection capacity, because data center flexibility can support the grid by delivering essential system-wide services like reserves and frequency regulation (e.g., aFRR). Finally, for regulators and policymakers, our data center flexibility and cost models can help assess the benefits of integrating data centers into power system flexibility markets.

This paper is a high-level comparative analysis of data center flexibility and cost. An important area for future work is the design of algorithms for real-time job scheduling to coordinate the real-time provision of data center flexibility services. In addition, this paper focuses on the temporal flexibility of data centers, that is, the flexibility of shifting computing workloads in time. Organizations with access to multiple data center sites also have spatial flexibility, which means the ability to shift workloads between locations. Evaluating spatial flexibility would require power flow analysis for the grids where the data centers are located.

\section{Acknowledgments}

This work was supported by the UK Engineering and Physical Sciences Research Council (EPSRC) (Project reference EP/S031901/1). Yihong Zhou's work was also supported by the Engineering Studentship from the University of Edinburgh; Ángel Paredes' work was supported by the FPU grant (FPU19/03791) founded by the Spanish Ministry of Education.

\newpage

\section{Methods} 
\label{sec:method}

This section is divided into several sub-sections for different parts of our results: 
\begin{enumerate}
    \item Section \nameref{method:max_size} describes the method for calculating the maximum amounts of flexibility. Related figures include Figs. \ref{fig:normalised_flex_AI}, \ref{fig:normalised_flex_General}, \ref{fig:all_HPC_analysis}, \ref{fig:all_HPC_corr}, and \ref{fig:size_vs_numdays}. 
    
    \item Section \nameref{method:HPC_cost} details the method for calculating the data center flexibility cost. Related figures include Figs. \ref{fig:unit_cost}, \ref{fig:interpret_cost_scaling_factor}, \ref{fig:unit_cost_lambda}, \ref{fig:all_HPC_analysis}, and \ref{fig:all_HPC_corr}. 
    
    \item Section \nameref{method:dq} describes the method for calculating the maximum flexibility and the cost with the dynamic quota opportunity (Fig. \ref{fig:flex_cost_dq}).
    
    \item Section \nameref{method:dataset_setting} describes the processing of the 14 HPC datasets in Table \ref{tab:summary}.
    
    \item Section \nameref{method:prob_settings} describes our case study settings.
    
\end{enumerate}

The following is a nomenclature for this section.

\nomenclature[F]{$\lvert \cdot \rvert$}{The cardinality function that returns the number of elements of a set.}
\nomenclature[F, 1]{$\lceil \cdot \rfloor$}{The round function that returns the nearest integer.}

\nomenclature[N, 1]{\(\mathcal{J}\)}{The index set of computing jobs.}
\nomenclature[N, 2]{\(\mathcal{T}\)}{The index set of time steps $\{1, \cdots, T\}$ in the optimization horizon.}
\nomenclature[N, 3]{\(\mathcal{T}_j\)}{The index set collecting the available time steps of job $j$.}
\nomenclature[N, 4]{\(\mathcal{T}^\text{A}_i\)}{The time steps to sustain flexibility over the $i^\text{th}$ activation of power system service.}
\nomenclature[N, 5]{\(\mathcal{N}^\text{A}\)}{The set that contains indices $i$ of the activation of power system service.}

\nomenclature[P, 1]{\( \delta^\text{max} \)}{The maximum allowable time delay (\%) of the job completion, relative to the computing time before providing flexibility.}
\nomenclature[P, 1]{\( t_j^\text{S} \)}{The submission time of job $j$.}
\nomenclature[P, 1]{\( t^\text{C}_j \)}{The completion time of job $j$ under the baseline scheduling strategy before providing flexibility.}
\nomenclature[P, 1]{\(M^\text{P}\)}{The time overhead triggered by each job preemption.}
\nomenclature[P, 1]{\(p^\text{base}_t\)}{The baseline data center electric power.}
\nomenclature[P, 2]{\(N^\text{R}_j\)}{The number of GPUs (respectively, CPUs) used by job $j$ specified by users in an AI-focused (respectively, general-focus) data center.}
\nomenclature[P, 3]{\(\hat{N}^\text{R}\)}{The total number of GPUs (respectively, CPUs) in an AI-focused (respectively, general-purpose) data center.}
\nomenclature[P, 4]{\( D_j \)}{The computing workload for job completion, which is the smallest number of time steps required to complete the job (computing time) under the user-specified number $N^\text{R}_j$ of computing resources.}
\nomenclature[P, 5]{\(G \)}{The increase of power (kW) of an AI-focused (respectively, general-purpose) data center when the utilization of a single GPU (respectively, CPU) increases from 0\% to 100\%, which can include the contribution from memory, cooling, etc. }
\nomenclature[P, 6]{\(G_0 \)}{The parameter that captures the fixed power (kW) independent of computing, such as the lighting power or the server idle power, of the whole data center.}
\nomenclature[P, 7]{$A_j/A$}{The price reduction coefficient.}
\nomenclature[P, 8]{$R$}{The hourly computing price of a single GPU/CPU.}
\nomenclature[P, 9]{$\pi$}{The energy price per kWh of electricity (USD/kWh).}
\nomenclature[P, 9]{$S$}{The amount (kW) of flexibility a data center aims to provide.}
\nomenclature[P, 9]{$K^\text{DQ}$}{The speed-up coefficient in the dynamic quota opportunity.}

\nomenclature[X, 1]{\( x_{j,t} \)}{The completed proportion of workload of job $j$ at time step $t$.}
\nomenclature[X, 2]{\( z_{j,t} \)}{The preemption counter variable that quantifies the amount of preemption for job $j$ at time step $t$.}
\nomenclature[X, 3]{\( n^P_j \)}{The total amount of job preemption of job $j$.}
\nomenclature[X, 4]{\( p_t \)}{The scheduled data center electric power.}
\nomenclature[X, 5]{\( f_t \)}{The flexibility at each time step $t$ in reducing demand.}
\nomenclature[X, 6]{\( s_i \)}{The amount of the provided flexibility over the $i^\text{th}$ activation of flexibility.}
\nomenclature[X, 7]{\( x_{j,t}' \)}{The binary variable that models the job running status at time step $t$.}
\nomenclature[X, 8]{\( e_j \)}{The scheduled completion time after flexibility provision.}
\nomenclature[X, 8]{\( \delta_j \)}{The delay proportion for job $j$.}
\nomenclature[X, 9]{\( C^\text{Flex} \)}{The total flexibility cost.}
\nomenclature[X, 9]{$t^\text{Ex}$}{The time delay of job $j$ after providing flexibility.}

\printnomenclature

\subsection{Solving the maximum amount of flexibility}
\label{method:max_size}

\subsubsection{Flexibility maximization problem}

The following optimization problem is solved to calculate the maximum amount of flexibility that a data center can deliver for a specific power system service.

\begin{maxi!}[3]
  {_{x_{j,t}, z_{j,t}, n_j^P, p_t, f_t, s_i}}{\frac{1}{|\mathcal{N}^\text{A}|}\sum_{i \in \mathcal{N}^\text{A}} s_i}{\label{eq:opt}}{\label{opt:obj}}
  \addConstraint{x_{j, t}}{\in [0, 1], z_{j, t} \in [0, 1], \quad}{\forall j\in \mathcal{J},\ t\in \mathcal{T} \label{opt:jobvar}}
  \addConstraint{x_{j, t}}{ = 0, z_{j, t} = 0, \quad}{\forall j\in \mathcal{J},\ t \notin \mathcal{T}_j \label{opt:jobzero}}
  \addConstraint{z_{j, t}}{\geq x_{j, t} - x_{j, t+1},\quad}{\forall j\in \mathcal{J},\ t\in \mathcal{T} \label{opt:anc_pausetime_cons}}
  \addConstraint{n^P_j}{ = \sum_t^{T} (z_{j, t}) - 1, \quad}{\forall j\in \mathcal{J} \label{opt:jobpausedelay_number1}}
  \addConstraint{\frac{M^\text{P}}{\Delta t}n^P_j}{\leq \epsilon \cdot D_j, \quad}{\forall j\in \mathcal{J} \label{opt:preemp_times}}
  \addConstraint{\sum_t^T x_{j, t}}{= D_j,\quad}{\forall j\in \mathcal{J} \label{opt:jobcompletion}}
  \addConstraint{\sum_j^J N^\text{R}_j \cdot x_{j, t}}{\leq \hat{N}^\text{R}, \quad}{\forall t\in \mathcal{T} \label{opt:GPUlim}}
  \addConstraint{p_t} {= \sum_j^J G \cdot N^\text{R}_j \cdot x_{j, t} + G_0,\quad}{\forall t\in \mathcal{T} \label{opt:powerrelation}}
  \addConstraint{f_t}{= p_t^\text{base} - p_t,}{\forall t\in \mathcal{T} \label{opt:flex_down}}
  \addConstraint{f_t}{\geq s_i,}{\forall t\in \mathcal{T}^\text{A}_i, \forall i \in \mathcal{N}^\text{A} \label{opt:flex_duration}}
  \addConstraint{s_i}{\geq 0,}{\forall i \in \mathcal{N}^\text{A} \label{opt:s}}
\end{maxi!}
The objective \eqref{opt:obj} is to maximize the amount of the power system service that the data center can provide on average over the $|\mathcal{N}^\text{A}|$  activation events. Constraints are listed below:


\eqref{opt:jobvar}: This constraint defines the value range for two decision variables. The continuous variable $x_{j, t}$ represents the completed workload of job $j$ at time step $t$. $x_{j, t} = 1$ represents the maximum workload achievable using the $N^\text{R}_j$ user-specified resources for a single time step. We assume that the computing workload per time step can be continuously adjusted. This is feasible under hardware control techniques such as dynamic power capping or DVFS applicable to both GPUs \cite{dutta2018gpu, Guerreiro-DVFS-modeling-for-energy-efficient-GPU-computing, krzywaniak2023dynamic} and CPUs \cite{krzywaniak2022depo, czarnul2019energy}. In addition, real-world data center schedulers (such as Run:AI) and the NVIDIA MIG functionality support fractional GPU allocation \cite{runai_gpu_partition, nvidia-mig}; CPUs are commonly allocated on a fractional basis in current data centers. Therefore, continuity may be better achieved by combining hardware controls and preemption of jobs with fractional resources. The continuous preemption counter variable $z_{j, t}$ is described in the explanation of \eqref{opt:anc_pausetime_cons}.


\eqref{opt:jobzero}: This constraint ensures that both $x_{j,t}$ and the preemption counter variable $z_{j,t}$ are zero outside the available period $\mathcal{T}_j$ of job $j$. The available period is defined as $\mathcal{T}_j \coloneqq \{t^S_j,\cdots, t^\text{C}_j + \lceil (1+\delta^\text{max})D_j\rfloor \}$, determined by the job submission time $t^S_j$ and the completion time $t^\text{C}_j$, incorporating the maximum proportion $\delta^\text{max}$ of job delay. 

\eqref{opt:anc_pausetime_cons}: This constraint models the the preemption counter variable $z_{j,t}$. As explained for \eqref{opt:jobvar}, when the completed workload $x_{j,t+1}$ decreases from the previous timestep $x_{j,t}$, it may be due to: 1) hardware control, and/or 2) preemption of jobs that use fractional resources. In the latter case, extra computing workload is required to save and reload checkpoints. $z_{j,t}$ considers the worst-case scenario and counts each workload decrease as preemption. The boundary condition $x_{j,T+1}=0$.

\eqref{opt:jobpausedelay_number1}: This constraint introduces $n^P_j$ to count the total amount of preemption for each job. The preemption counter $z_{j, t}$ counts the decrease of $x_{j, t}$ for job completion, which is not considered preemption and is thus subtracted from $n^P_j$.

\eqref{opt:preemp_times}: This constraint ensures that the total extra workloads due to preemption should be minor compared to the computing workload $D_j$ for job completion. Here, $\epsilon$ is a small positive number. $M^\text{P}$ (minutes) is the extra computing workload per preemption. $\Delta t$ is the length (minutes) of a single time step.

\eqref{opt:jobcompletion}: This constraint ensures job completion. Completing job $j$ requires a computing workload $D_j$, which is the smallest number of time steps (computing time) required to complete the job using the user-specified number ($N^\text{R}_j$) of computing resources. As mentioned in Section \nameref{res:size_of_flex} (in \nameref{sec:results}), the computing resources refer to CPUs for general-purpose HPC data centers and GPUs for AI-focused HPC data centers.

\eqref{opt:GPUlim}: This constraint ensures that the total used resources at each time step do not exceed the total number ($\hat{N}^\text{R}$) of resources in the data center. 


\eqref{opt:powerrelation}: This constraint establishes a linear relationship between the data center electric power $p_t$ and the computing workload $x_{j, t}$. The coefficient $G$ in kW represents the increase in power when the utilization of a single unit of computing resource increases from 0\% to 100\%, which can include the contribution of memory, cooling, etc. $G_0$ in kW captures fixed power demand which is independent of computing (e.g. from lighting or idle power). 



\eqref{opt:flex_down}: This constraint models the demand-reduction flexibility $f_t$. The baseline power $p_t^\text{base}$ is a parameter and is inferred from the data center datasets using the same power model \eqref{opt:powerrelation}.

\eqref{opt:flex_duration}: This constrains that the flexibility must be sustained at a specific amount $s_i$ for each service activation indexed by $i\in \mathcal{N}^\text{A}$, where the $i^\text{th}$ activation spans several consecutive time steps collected in the set $\mathcal{T}^\text{A}_i$. The amount of flexibility $s_i$ must be non-negative as specified in \eqref{opt:s}. Different types of power system service have different frequency $|\mathcal{N}^\text{A}|$ and duration $|\mathcal{T}^\text{A}_i|$ requirements, where $|\cdot|$ calculates the number of elements in the set. For a specific type of power system service, the duration of each activation is the same: $|\mathcal{T}^\text{A}_i| = |\mathcal{T}^\text{A}_k|, \forall i,k \in \mathcal{N}^\text{A}$. 



\subsubsection{Scaling the maximum amounts of flexibility}
\label{method:scale_size}


We solve \eqref{eq:opt} to get the maximum flexibility under a set of nominal data center parameters. Due to the use of linear power model in \eqref{opt:powerrelation}, for data centers with other parameter settings, we can directly scale the maximum flexibility under the nominal parameters without time-consuming re-optimization. We use overhead lines $\overline{\blacksquare}$ to specify our nominal parameters. It is worth noting that we must have a specific nominal setting $\overline{G_0}=0$ for the following derivation. The derived scaling formula can scale the result for arbitrary settings of $G_0$. 



Assuming all decision variables take the values of the optimal solution, the maximum amount $S$ of flexibility by solving \eqref{eq:opt} can be expressed as:
\begin{equation}
    S =  ave_{i\in \mathcal{N}^\text{A}} \min_{t\in \mathcal{T}_i^A} f_t
\end{equation}
The operator $\min_{t\in \mathcal{T}_i^A}(\cdot)$ finds the sustained amount of flexibility during the $i^\text{th}$ activation. Based on Eqs. \eqref{opt:flex_down} and \eqref{opt:powerrelation}, $f_t$ can be expressed as:
\begin{flalign}
    \nonumber  f_t & =  p^\text{base}_t - p_t & \\
    \nonumber & = \sum_j^J G \cdot N^\text{R}_j \cdot x^\text{base}_{j, t} + G_0 - \sum_j^J G \cdot N^\text{R}_j \cdot x_{j, t}-G_0 \hspace{-10pt} &\\
     & = \sum_j^J G \cdot N^\text{R}_j \cdot (x^\text{base}_{j, t} - x_{j, t}) &
\end{flalign}
where the superscript $\blacksquare^\text{base}$ denotes values corresponding to the baseline utilization of the data center (before flexibility provision). Then, the maximum amount of flexibility can be re-written as:
\begin{equation}
    S =  ave_{i\in \mathcal{N}^\text{A}} \min_{t\in \mathcal{T}_i^A} \left\{ \sum_j^J G \cdot N^\text{R}_j \cdot (x^\text{base}_{j, t} - x_{j, t}) \right\}
\label{eq:max_size_rewritten}
\end{equation}

Figs. \ref{fig:normalised_flex_AI}, \ref{fig:normalised_flex_General}, \ref{fig:all_HPC_analysis}, \ref{fig:all_HPC_corr}, \ref{fig:normalised_flex_AI_dq0.5}, and \ref{fig:normalised_flex_General_dq0.5} show the normalized maximum amounts of flexibility, which is the flexibility in kW divided by the data center maximum power. Here, the maximum power $p^\text{max}$ is defined as the power when all computing devices are used 100\%. In other words:
\begin{equation}
    p^\text{max} = G \cdot \hat{N}^\text{R} + G_0
\label{eq:pmax}
\end{equation}
The normalized flexibility $S^\text{norm}$ is therefore expressed as:
\begin{equation}
\begin{split}
    & S^{\text{norm}} = \frac{S}{p^\text{max}} \\
    & = ave_{i\in \mathcal{N}^\text{A}} \min_{t\in \mathcal{T}_i^A} \left\{
    \frac{\sum_j^J G \cdot N^\text{R}_j \cdot (x^\text{base}_{j, t} - x_{j, t})}{G \cdot \hat{N}^\text{R} + G_0}
    \right\}
\end{split}
\label{eq:max_size_norm}
\end{equation}
Then, the normalized flexibility under our nominal parameters (recall that $\overline{G_0}=0$) can be expressed as:
\begin{equation}
    \overline{S^\text{norm}} = 
    ave_{i\in \mathcal{N}^\text{A}} \min_{t\in \mathcal{T}_i^A} \left\{
    \frac{\sum_j^J \cdot \overline{N^\text{R}_j} \cdot (x^\text{base}_{j, t} - x_{j, t})}{\overline{\hat{N}^\text{R}}}
    \right\}
\label{eq:max_size_norm_own}
\end{equation}
Here we assume other data centers have similar job characteristics and the only differences are parameters $N^\text{R}_j$, $\hat{N}^\text{R}$, $G$ and $G_0$. In other words, we assume that jobs in other data centers utilize the same proportion of the total data center resources:
\begin{equation}
    \frac{\overline{N^\text{R}_j}}{N^\text{R}_j} = \frac{\overline{\hat{N}^\text{R}}}{\hat{N}^\text{R}}, \forall j\in \mathcal{J}
\label{eq:assump}
\end{equation}
Combining Eqs. \eqref{eq:max_size_norm}, \eqref{eq:max_size_norm_own}, and \eqref{eq:assump}, the normalized amount of flexibility $S^\text{norm}$ under arbitrary parameters $N^\text{R}_j$, $\hat{N}^\text{R}$, $G$ and $G_0$ can be calculated by a linear operation of the normalized flexibility under our parameter setting $\overline{S^\text{norm}}$:

\begin{equation}
    S^{\text{norm}} = \overline{S^\text{norm}} \cdot \frac{G \cdot \hat{N}^\text{R}}{G \cdot \hat{N}^\text{R} + G_0}
\label{eq:scale_size_norm}
\end{equation}
By combining \eqref{eq:pmax}, the amount of flexibility $S$ in kW after scaling is:
\begin{equation}
    S= S^{\text{norm}} \times p^\text{max} = \overline{S^\text{norm}} \cdot G \cdot \hat{N}^\text{R}
\label{eq:scale_size}
\end{equation}
Eq. \eqref{eq:scale_size_norm} suggests that the normalized flexibility results in our figures (such as \ref{fig:normalised_flex_AI} and \ref{fig:normalised_flex_General}) are independent of data center parameters, as long as there is negligible fixed power that is independent of computing (i.e., $G_0=0$).


It is important to note that, while the assumption stated in \eqref{eq:assump} could produce imprecise scaling results if other data centers exhibit different job characteristics, our scaling formula can still help other data centers to quickly evaluate their flexibility by scaling the results calculated in our nominal setting (or via our interactive Google Colab page \cite{dace_api}).

\subsection{Data center cost of flexibility} 
\label{method:HPC_cost}

\subsubsection{Cost modeling}
\label{method:cost_interp}

As explained in Section \nameref{res:cost_flex} (in \nameref{sec:results}), The total cost for the data center providing a particular power system service is the sum of the price reduction, which is incurred by the delays of jobs. We propose using the following linear model to express the price reduction $c_j$ for each job $j$:
\begin{equation}
     c_j = A_j \cdot \delta_j \cdot V_j
\label{eq:comp_cost_def}
\end{equation}
where $V_j$ is the original computing price of the job, $A_j$ is the price reduction coefficient, measuring the proportionate price reduction in response to a certain delay proportion $\delta_j$. The delay proportion $\delta_j$ is defined as the ratio of the extra time $t_j^{Ex}$ beyond the original job completion time relative to the computing time $D_j$:
\begin{equation}
    \delta_j = \frac{t^\text{Ex}_j}{D_j}
\label{eq:delay_prop}
\end{equation}
In several cloud computing platforms such as Google Cloud, AWS, Oracle, and Lambda Cloud, the computing price $V_j$ is calculated as:
\begin{equation}
    V_j = D_j \cdot R \cdot N^\text{R}_j
\label{eq:value}
\end{equation}
where $R$ is the hourly price of a single unit of computing resource and $N^\text{R}_j$ is the user-specified number of resources. Recall that the computing resources refer to GPUs for AI-focused HPC data centers and refer to CPUs for general-purpose HPC data centers. Also note that most cloud GPU prices will include necessary CPU, memory, and storage.

As explained above, the total flexibility cost $C^\text{Flex}$ is:
\begin{equation}
    C^\text{Flex} = \sum_{j\in \mathcal{J}} c_j
\label{eq:tot_price_reduction}
\end{equation}

\subsubsection{Cost minimization problem}
\label{method:cmin}

The following problem is solved to find the data center cost of providing $S$ amount of flexibility for a specific power system service.
\begin{mini!}[3]
  {\substack{x_{j,t}, x_{j,t}', z_{j,t}, n_j^P, p_t, f_t, s_i, e_j, \delta_j, c_j}}{C^\text{Flex}}{\label{cmin}}{\label{cmin:obj}}
  \addConstraint{\eqref{opt:jobvar}-\eqref{opt:s}}{}{}
  \addConstraint{x_{j,t}'}{\geq x_{j,t},\quad}{\forall j\in \mathcal{J},\ t\in \mathcal{T} \label{opt:running_indicator}}
  \addConstraint{e_j}{\geq t\cdot x_{j,t}' + 1, \quad}{\forall t \geq t^S_j + D_j, \forall j \in \mathcal{J} \label{cmin:complete_t}}
  \addConstraint{\delta_j}{\geq 0, \delta_j \geq \frac{e_j - t^S_j - D_j}{D_j},}{ \forall j \in \mathcal{J} \label{cmin:delay_prop}}
  \addConstraint{c_j}{\geq A_j \cdot \delta_j \cdot  D_j \cdot R \cdot N^\text{R}_j, \quad}{\forall j \in \mathcal{J} \label{cmin:comp_cost}}
  \addConstraint{\frac{1}{|\mathcal{N}^\text{A}|}\sum_{i \in \mathcal{N}^\text{A}} s_i}{\geq S}{\label{cmin:min_size}}
\end{mini!}
Constraint \eqref{opt:running_indicator} introduces binary variables $x_{j,t}'$ to indicate the job running status. In \eqref{cmin:complete_t}, the decision variable $e_j$ models the last time step of job running. Therefore, the term $e_j - t^S_j - D_j$ in \eqref{cmin:delay_prop} represents the time delay $t^\text{Ex}$. Constraint \eqref{cmin:comp_cost} models the reduction in computing price as defined in \eqref{eq:comp_cost_def}. Constraint \eqref{cmin:min_size} ensures that the average flexibility over the $|\mathcal{N}^\text{A}|$ times of service activation is not less than the specified $S$.

After solving \eqref{cmin}, we calculate the average cost of flexibility ($ACoF$) by dividing $C^\text{Flex}$ by the total shifted energy over the service activation periods. $ACoF$ is exactly the average flexibility cost in all our cost plots, such as in Fig. \ref{fig:unit_cost}.


\subsubsection{Cost minimization problem tightening}
\label{method:cmin_tight}

The cost minimization problem \eqref{cmin} is a challenging mixed-integer linear programming (MILP). To speed up the computation, we introduce a tightening method by integrating a valid lower bound of the cost into \eqref{cmin}. This method assumes that there is no queuing time, meaning submission time equals start time.

Based on Eqs. \eqref{eq:comp_cost_def} and \eqref{eq:delay_prop}, the reduction of the computing price for job $j$ can be rewritten as:
\begin{equation}
    c_j = A_j \cdot \delta_j \cdot  D_j \cdot R \cdot N^\text{R}_j = A_j \cdot R \cdot t^\text{Ex}_j \cdot N^\text{R}_j
\label{eq:tot_cost_re}
\end{equation}
For subsequent derivation, we need to assume that all jobs have the same price reduction coefficient $A = A_j$. In practice, the value of $A$ can be calculated as a weighted average of the coefficients of $A_j$ for all jobs. Under this assumption, the total flexibility cost $C^\text{Flex}$ becomes:
\begin{equation}
    C^\text{Flex} = \sum_{j \in \mathcal{J}} c_j = A (\sum_{j \in \mathcal{J}} t^\text{Ex}_j \cdot N^\text{R}_j) \cdot R
\end{equation}
Because $t^\text{Ex}_j$ is the extra time beyond the original job completion time and $N^\text{R}_j$ is the number of computing resources, the term $(\sum_{j \in \mathcal{J}} t^\text{Ex}_j \cdot N^\text{R}_j)$ reflects the effort to complete the ``total delayed computing workload of all jobs $W^\text{delay}$''. Furthermore, $(\sum_{j \in \mathcal{J}} t^\text{Ex}_j \cdot N^\text{R}_j)$ is an upper bound of $W^\text{delay}$, because jobs over their delayed times $t^\text{Ex}_j$ may not 100\% utilize their computing resources due to the data center capacity limit. In other words:
\begin{equation}
    W^\text{delay} \leq (\sum_{j \in \mathcal{J}} t^\text{Ex}_j \cdot N^\text{R}_j)
\label{eq:lb}
\end{equation}
When there is no dynamic quota, providing flexibility will definitely trigger workload delays. Based on our linear power model \eqref{opt:powerrelation}, the delayed computing workload is proportional to the shifted energy over flexibility provision through the power coefficient $G$. Now, consider the provision of $S$ units of flexibility for $|\mathcal{N}^\text{A}|$ times with each of the flexibility activation sustaining $|\mathcal{T}^\text{A}_i|$ time steps, the total shifted energy is $|\mathcal{T}^\text{A}_i| \cdot |\mathcal{N}^\text{A}| \cdot S$, so the ``total delayed computing workload of all jobs $W^\text{delay}$'' can be expressed as:
\begin{equation}
    W^\text{delay} = |\mathcal{T}^\text{A}_i| \cdot |\mathcal{N}^\text{A}| \cdot \frac{S}{G}
\end{equation}
which, combined with \eqref{eq:tot_cost_re} and \eqref{eq:lb}, leads to a lower bound $\underline{C}$ of the total flexibility cost $C^\text{Flex}$, namely the total reduction in computing prices:
\begin{equation}
\label{eq:comp_lb}
     \underline{C} =  A \cdot |\mathcal{T}^\text{A}_i| \cdot |\mathcal{N}^\text{A}| \cdot R \cdot \frac{S}{G} \leq C^\text{Flex} = \sum_{j \in \mathcal{J}} c_j
\end{equation}
In our simulation, setting \eqref{eq:comp_lb} as an additional constraint in \eqref{cmin} reduces the solution time significantly.

\subsubsection{Scaling the cost of flexibility provision}
\label{method:scale_cost}

We solve \eqref{cmin} to get the minimum flexibility cost $C^\text{Flex}$ and then the average flexibility cost $ACoF$ under a set of nominal data center parameters. Due to the use of linear cost model, for other data center parameters, we can directly scale the $ACoF$ calculated under the nominal parameters without re-optimization. We use overhead lines $\overline{\blacksquare}$ to specify our nominal parameters.

Again, we assume that all decision variables take the values of the optimal solution. Note that we have assumed $A_j = A$ in Section \nameref{method:cmin_tight}. Then based on Eq. \eqref{eq:comp_cost_def}, the (minimum) total cost of providing a certain amount $S$ of flexibility can be expressed as:
\begin{equation}
    C^\text{Flex} = \sum_{j\in \mathcal{J}} c_j = A \cdot \sum_{j\in \mathcal{J}} \delta_j \cdot  D_j \cdot R \cdot N^\text{R}_j
\label{eq:tot_comp_cost}
\end{equation}
The total shifted energy $E^f$ in kWh over the service activation periods can be expressed as:
\begin{equation}
    E^f \coloneqq \Delta t \cdot |\mathcal{N}^\text{A}| \cdot |\mathcal{T}^\text{A}_i| \cdot S
\end{equation}
Based on Eq. \eqref{eq:scale_size}, we further have
\begin{equation}
    E^f = \Delta t \cdot |\mathcal{N}^\text{A}| \cdot |\mathcal{T}^\text{A}_i| \cdot \overline{S^\text{norm}} \cdot \hat{N}^\text{R} \cdot G
\label{eq:tot_e_cost}
\end{equation}
Note that, due to the use of \eqref{eq:scale_size}, we need to set the nominal parameter $\overline{G_0}=0$ to maintain the initial assumption. Given \eqref{eq:tot_comp_cost} and \eqref{eq:tot_e_cost}, we have: 
\begin{equation}
    ACoF = \frac{C^\text{Flex}}{E^f} = \frac{A \cdot \sum_{j\in \mathcal{J}} \delta_j \cdot  D_j \cdot R \cdot N^\text{R}_j}{\Delta t \cdot |\mathcal{N}^\text{A}| \cdot |\mathcal{T}^\text{A}_i| \cdot \overline{S^\text{norm}} \cdot \hat{N}^\text{R} \cdot G}
\label{eq:ADCoF}
\end{equation}
Again, we assume that if \eqref{eq:assump} holds, the average flexibility cost under our nominal parameter setting can be expressed as:
\begin{equation}
    \overline{ACoF} = \frac{\overline{A} \cdot \sum_{j\in \mathcal{J}} \delta_j \cdot  D_j \cdot \overline{R} \cdot N^\text{R}_j}{\Delta t \cdot |\mathcal{N}^\text{A}| \cdot |\mathcal{T}^\text{A}_i| \cdot \overline{S^\text{norm}} \cdot \hat{N}^\text{R} \cdot \overline{G}}
\label{eq:ADCoF_own}
\end{equation}
Combining \eqref{eq:ADCoF} and \eqref{eq:ADCoF_own}, given $\overline{ACoF}$ calculated in our nominal setting, the formula to scale the $ACoF$ according to other data center parameters $A$, $R$, and $G$ is:
\begin{equation}
    ACoF = \overline{ACoF} \cdot (A\cdot R \cdot \overline{G})/(G \cdot \overline{A} \cdot \overline{R})
\label{eq:scale_ACoF_nodq}
\end{equation}
The term $(A\cdot R \cdot \overline{G})/(G \cdot \overline{A} \cdot \overline{R})$ is the \textit{cost scaling factor}. 


\subsubsection{Estimating the cost scaling factor}

The sections above provide the method for calculating the flexibility cost in a nominal setting (can be set freely with $\overline{G_0}=0$), and describe how to scale the cost to other parameter settings with the cost scaling factor. However, a question remains of how to estimate a cost scaling factor that allows us to adjust the nominal cost result to reflect the flexibility costs of real-world data centers. Answering this question is the foundation of all our cost plots, such as Figs. \ref{fig:unit_cost}, \ref{fig:interpret_cost_scaling_factor}, \ref{fig:unit_cost_lambda}, \ref{fig:all_HPC_analysis}, and \ref{fig:all_HPC_corr}. This section describes the estimation method based on real-world cloud platform data.

Recall that our flexibility cost is based on the assumption that slower computing will lead to a lower computing price. Cloud platforms provide several CPU and GPU rental options with different computing speeds and prices, based on which we can estimate the cost scaling factor.

The method is illustrated by the following example. Suppose a cloud platform provides two computing rental options. Option I has $N^R_1$ computing resources, a computing speed of $O_1$, hourly price per computing resource of $R_1$, and the power per computing resource is $G_1$ (in kW), while the parameters for Option II are $N^R_2$, $O_2$, $R_2$, and $G_2$. Suppose a computing job has computing workload of $W$. Then, if Option I is rented, the computing time is $D_1 = W/O_1$. Based on \eqref{eq:value}, the computing price is $V_1 = W/O_1 \cdot R_1 \cdot N^R_1$. Similarly, if Option II is rented, the computing time is $D_2 = W/O_2$, and the price is $V_2 = W/O_2 \cdot R_2 \cdot N^R_2$. Without loss of generality, we assume that Option I is the faster and more expensive option, namely $D_1 < D_2$ and $V_1 > V_2$. Now, if Option I is used to provide power system flexibility service and the computing time is increased to $D_1'=D_2$ (job delay), then the computing price of the delayed Option I should not exceed the price of Option II. In other words, the flexibility cost of Option I is:
\begin{equation}
    c_1 = V_1 - V_2
\end{equation}
By definition, the delay proportion of Option I for providing flexibility is $\delta_1 = (D_2-D_1)/D_1$. Then, based on Eq. \eqref{eq:comp_cost_def}, we can estimate the price reduction coefficient of Option I as:
\begin{equation}
    A_1 = c_1 / (\delta_1 \cdot V_1)
\end{equation}
Then, based on \eqref{eq:scale_ACoF_nodq}, we can estimate the corresponding cost scaling factor as $(A\cdot R \cdot \overline{G})/(G \cdot \overline{A} \cdot \overline{R})$.

Through the example above, we see that we can get the cost scaling factor by comparing two computing rental options. Therefore, we collect all GPU and CPU rental options available in Google Cloud, AWS, Oracle, and Lambda Cloud. It should be noted that cloud platforms provide CPUs not just for HPC but also for other applications such as web services. As we focus on HPC data centers, we only select a subset of CPU options on cloud platforms with optimized computing performance. Details of the collection process can be found in our supplementary file. The CPU speed information can be found on the PassMark website \cite{passmark}. We use the CPU mark as the speed information, which is the weighted harmonic average of the computing speed for several benchmark tests (e.g., integer math, floating point math, physics) that are important for HPC applications \cite{passmark_scoremean}. The GPU speed information is available at Lambda Lab \cite{lambda_gpu_benchmark}, which is the weighted average speed for a range of AI tasks. The speed information is provided on both the FP32 and FP16 basis, and we use the harmonic average of the two as the final GPU speed. More than 100 CPU/GPU options with the complete desired information are recorded. We set $G$ as the device rated power. Note that, as the utilization of a device (GPU or CPU) increases from 0 to 100\%, the power consumption may be lower than the device rated power due to the idle power, leading to an overestimation of $G$. However, there are also power increases in memory and cooling that could approximately counterbalance the overestimation of $G$. Finally, on most cloud platforms, CPUs are rented on the basis of virtual CPUs (vCPUs), which typically represent one thread of a physical CPU. The power of a vCPU is calculated by dividing the rated power of the physical CPU by the number of vCPUs.

With more than 100 CPU/GPU options from these cloud service providers, for either CPU (represents general-purpose data centers) or GPU (represents AI-focused data centers), we compare any two options within the same provider to estimate the corresponding cost scaling factor, as illustrated in the previous example. We avoid comparing options across different platforms because their additional services also affect the pricing. We do not compare the computing options if one would require 100\% more computing time than the other. This is to exclude options that are excessively slow and therefore less comparable. In addition, there are comparisons where one option is slower and more expensive than the other, and these comparisons are dropped. These ``slower and more expensive'' options exist because (potentially) there is a shortage in other computing resources or they have larger memory or storage spaces. 

Note that, as mentioned in the main content, the Lambda Cloud GPU prices are significantly lower than the other three cloud platforms. As in Fig. \ref{fig:unit_cost_lambda}, the analysis of Lambda Cloud's data has been separated from the other three cloud platforms.

\subsection{Dynamic quota}
\label{method:dq}

\subsubsection{Solving the maximum flexibility under dynamic quota}

To model the dynamic quota functionality, we introduce another set of continuous variables $x^\text{DQ}_{j, t} \in [0, 1]$ and a parameter $K^\text{DQ} \in [0, 1]$ that represents the speed-up coefficient. In other words, using 100\% more computing resources will complete $K^\text{DQ}$ more computing workload in a single time step. The problems to find the maximum amount of flexibility with dynamic quota have the following set of updated constraints:
\begin{itemize}
    \item \eqref{opt:jobcompletion} is updated to
    \begin{equation}
        \sum_t^T (x_{j, t} + K^\text{DQ} x^\text{DQ}_{j,t}) = D_j
    \end{equation}

    \item \eqref{opt:GPUlim} is updated to
    \begin{equation}
        \sum_{j}^J N^\text{R}_j \cdot (x_{j, t} + x^\text{DQ}_{j, t}) \leq \hat{N}^\text{R}
    \end{equation}

    \item \eqref{opt:powerrelation} is updated to
    \begin{equation}
        p_t = \sum_j^J G \cdot N^\text{R}_j \cdot (x_{j, t} + x^\text{DQ}_{j, t}) + G_0
    \end{equation}

    \item Adding more computing resources may not always result in speed-up due to communication delays across devices. To avoid overestimating the capability of dynamic quota, we limit the number of additional resources to not exceed the originally specified resources through:
    \begin{equation}
        x^\text{DQ}_{j,t} \leq x_{j,t}
    \end{equation}
\end{itemize}

\subsubsection{Calculating the flexibility cost under dynamic quota}

As extra computing resources typically do not achieve a full 100\% speed-up, activating the dynamic quota would result in additional energy costs, alongside the flexibility cost triggered by the reduction in computing prices as defined in \eqref{eq:tot_price_reduction}. The extra energy cost can be expressed as: 
\begin{equation}
\label{eq:extra_energy_cost}
    C^\text{E} = \pi \Delta t(\sum_{t\in \mathcal{T}} p_t - \sum_{t\in \mathcal{T}} p^\text{base}_t)
\end{equation}
where $\pi$ is the energy price and $\Delta t(\sum_{t\in \mathcal{T}} p_t - \sum_{t\in \mathcal{T}} p^\text{base}_t)$ is the additional energy consumption. The total flexibility cost with the dynamic quota opportunity becomes:
\begin{equation}
\label{eq:cost_def_dq}
    C^\text{Flex} = \sum_{j\in \mathcal{J}} c_j + C^\text{E}
\end{equation}
Eqs. \eqref{eq:extra_energy_cost} and \eqref{eq:cost_def_dq} are included in the minimization problem \eqref{cmin} to calculate the flexibility cost under dynamic quota.

\subsubsection{Cost minimization problem tightening under dynamic quota}

Because job delays could be avoided by using spare resources to accelerate computation, the lower bound in \eqref{eq:comp_lb} is not valid for dynamic quota, and an updated lower bound is required. Suppose $S_a$ is the maximum flexibility without causing any job delays in the dynamic quota opportunity. In case where delivered flexibility $S_b > S_a$, the new valid lower bound for the total computing price reduction is:
\begin{equation}
     \underline{C} =  A \cdot |\mathcal{T}^\text{A}_i| \cdot |\mathcal{N}^\text{A}| \cdot R \cdot \frac{S_b - S_a}{G(1+K^\text{DQ})}
\end{equation}
To see this, the profile offering the maximum flexibility $S_a$ under zero job delay can be considered a new baseline profile. Then, providing $(S_b - S_a)$ additional flexibility will certainly trigger job delays and thus the reduction in computing prices. Therefore, the lower bound of the total price reduction can be obtained by substituting $S$ with $(S_b - S_a)$ in \eqref{eq:comp_lb}, which is further divided by $1+K^\text{DQ}$ that represents the computing speed-up with dynamic quota.

\subsubsection{Cost scaling under dynamic quota}

With the dynamic quota opportunity, there will be extra energy cost, so the $ACoF$ would additionally include the average extra energy cost of flexibility ($AECoF$). To differentiate, here we refer to the $ACoF$ without dynamic quota (Eq. \eqref{eq:scale_ACoF_nodq}) as $APCoF$. The $ACoF$ with the dynamic quota opportunity is the sum of $APCoF$ and $AECoF$. The scaling of $APCoF$ still follows \eqref{eq:scale_ACoF_nodq}. As for $AECoF$, first, based on Eq. \eqref{opt:powerrelation}, the extra energy cost $C^\text{E}$ in \eqref{eq:extra_energy_cost} can be re-express as:
\begin{equation}
    C^\text{E} = \pi \Delta t\sum_{t\in \mathcal{T}} \sum_{j\in \mathcal{J}} G \cdot N^\text{R}_j \cdot (x^\text{base}_{j, t} - x_{j, t})
\end{equation}
Combined with Eq.\ \eqref{eq:tot_e_cost}, $AECoF$ can be expressed as:
\begin{equation}
    AECoF = \frac{C^\text{E}}{E^f} = \frac{\pi \Delta t\sum_{t\in \mathcal{T}} \sum_j^J  \cdot N^\text{R}_j \cdot (x^\text{base}_{j, t} - x_{j, t})}{\Delta t \cdot |\mathcal{N}^\text{A}| \cdot |\mathcal{T}^\text{A}_i| \cdot \overline{S^\text{norm}} \cdot \hat{N}^\text{R} } 
\label{eq:aecof}
\end{equation}
Again we assume that if \eqref{eq:assump} holds, following a similar process, given the $\overline{AECoF}$ calculated in our nominal parameter $\overline{\pi}$, the formula to scale to other parameter settings is:
\begin{equation}
    AECoF = \overline{AECoF} \cdot \frac{\pi}{\overline{\pi}}
 \label{eq:scale_AECoF}
\end{equation}
Therefore, the scaling formula for $ACoF$ with the dynamic quota opportunity is:
\begin{equation}
\begin{split}
    ACoF & = APCoF + AECoF \\
    & = \overline{APCoF} \cdot \frac{A\cdot R \cdot \overline{G}}{G \cdot \overline{A} \cdot \overline{R}} + \overline{AECoF} \cdot \frac{\pi}{\overline{\pi}}
\end{split}
\label{eq:scale_ACoF}
\end{equation}

\subsection{Data processing}
\label{method:dataset_setting}

This section presents data processing of all 14 HPC data center datasets used in our case studies. Some background information is provided. There are three important time steps for each job record: the submission time, start time, and completion time. The submission time is when the user commits the job. The start time is when the submitted job begins running, which may be later than the submission time due to resource unavailability. The completion time is when the job is completed. The difference between the completion time and the start time is the job computing time. Because some of our data center datasets lack job submission times, for consistent comparisons, we set all submission times equal to the start times, resulting in zero queue time. This can further lead to an underestimation of the data center flexibility, since there could have been queue-time flexibility.

\subsubsection{Data selection}

Among the 14 HPC data center datasets, some of them contain job data that span several years. To reduce simulation complexity and ensure up-to-dateness, we selected 80 consecutive days of job records from the most recent year. Note that, the data center recorder might miss jobs that were submitted before or completed after the recording period, which leads to under-utilization at the beginning or the end of the recording period. These under-utilized periods are trimmed accordingly. Some datasets contain fewer than 80 days of records post-trimming, thus only 40 or 60 consecutive days are chosen. 

\subsubsection{Short-duration jobs}
\label{method:short_dur_jobs}

Job start times and completion times are rounded to our time resolution $\Delta t$. For jobs having zero computing time after rounding, we set their computing times to one time step and rescale the required number of resources to ensure the same computing workload: the product of the computing time and the required number of resources.

\subsubsection{Job partition}
\label{method:job_partition}

For jobs started before and/or completed after the optimization horizon $\mathcal{T}$, we truncate their computing times to equal the number of time steps of job running within $\mathcal{T}$ under the baseline profiles.

For example, suppose our optimization horizon is 00:00 Day 1 - 24:00 Day 1, if there is a job started at 18:00 Day 0 (6 hours before the horizon) and completed at 2:00 Day 2 (2 hours after the horizon) in the baseline profile, then we set the computing time of this job to 24 hours.

As the second example, considering the same optimization horizon, if there is a job started at 18:00 Day 1 and completed at 2:00 Day 2 (2 hours after the horizon), in this case, this job only needs computation for 24:00D1 - 18:00D1 = 6 hours.

\begin{figure}[tb]
    \centering
    \includegraphics[width=1\linewidth]{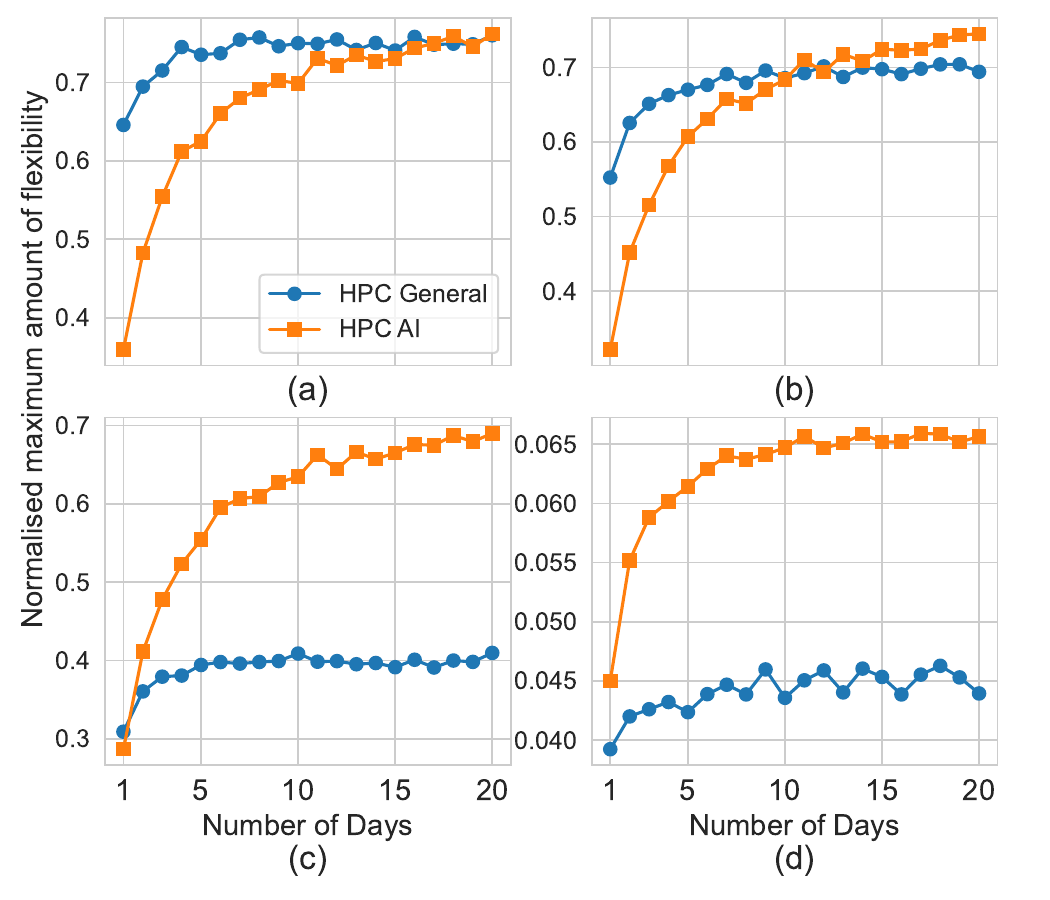}
    \vspace{-9 mm}
    \caption{The maximum amounts of flexibility calculated by solving optimization problems with varying horizons (number of days). The calculation is based on the two HPC data centers in Fig. \ref{fig:uti_base_max_amount}. (a) Results for a power system service with duration of 0.25 hours and frequency of 365 times/year. (b) Results when the duration is 0.25 hours and the frequency is 2,920 times per year. (c) Results when the duration is 2 hours and the frequency is 365 times per year. (d) Results when the duration is 2 hours and the frequency is 2,920 times per year. The increasing trend of the maximum amount of flexibility with the increasing optimization horizon confirms our anticipation of reduced conservativeness. This figure shows that the AI-focused HPC data center in Figs. \ref{fig:normalised_flex_AI} and \ref{fig:normalised_flex_General} should have maximum flexibility similar to the general-purpose HPC data center. This holds true even for short-duration, low-frequency services when the optimization horizon increases beyond 15 days.}
    \label{fig:size_vs_numdays}
    \vspace{-2 mm}
\end{figure}

This partition strategy can lead to conservativeness in finding the maximum flexibility, as jobs completed after the optimization horizon will have no ability to delay (as in the two examples above). This conservativeness increases for data centers with more long-computing-time jobs, such as the AI-focused HPC data center, as previously observed in Fig. \ref{fig:normalised_flex_AI}. This conservativeness will reduce with the increase of the optimization horizon, as illustrated in Fig. \ref{fig:size_vs_numdays}. Despite of the conservativeness, an advantage of this partition strategy is the feasibility guarantee for the next-horizon problem, since this problem does not carry out more computing workload than that before flexibility provision. This also means that optimization problems under non-overlapping horizons are independent, so these problems for different date ranges can be solved in parallel. One may consider another partition strategy such that jobs only need to complete a portion of the computing workload within the current optimization horizon. However, this may lead to an overestimate of the flexibility cost because some jobs are forced to be delayed unnecessarily.

\subsubsection{Job aggregation}
\label{method:job_agg}

Some of our 14 HPC datasets contain numerous jobs that are running simultaneously, which leads to complicated optimization problems used to find the maximum amount of flexibility \eqref{eq:opt} and the minimum flexibility cost \eqref{cmin}. Since this paper focuses on high-level evaluation rather than production-level implementation, we use a job aggregation strategy to reduce the computational complexity.

We aggregate daily jobs into 100 groups using K-means clustering based on the start time and completion time, then replace each group with a single aggregated job, resulting in only 100 jobs per day after aggregation. For the aggregated job of each group, the start time and the submission time are the earliest start and submission times of the original jobs in the group, and the completion time is the latest. The computing time of the aggregated job is the difference between the completion and start times. The number of resources (GPUs/CPUs) used for each aggregated job is calculated by summing the products of computing time and the number of resource for each job $j$ in the group, then dividing by the computing time of the aggregated job. This process ensures that the computing workload of the aggregated job is equal to the total workloads of the original jobs in that group. 

The aggregation strategy improves computational efficiency while preserving utilization fidelity. Our supplementary file plots the utilization curves based on the aggregated jobs and those based on the original jobs. 

\subsection{Problem settings}
\label{method:prob_settings}

The optimization horizon is set to 10 days with a resolution $\Delta t$ of 15 minutes, resulting in $T=960$. 
In the 10-day optimization horizon, we randomly pick $|\mathcal{N}^\text{A}|$ non-overlapping flexibility activation periods from a uniform distribution, where each period contains $|\mathcal{T}^\text{A}_i|$ consecutive time steps. This achieves an annual flexibility frequency of $365 / 10 \times |\mathcal{N}^\text{A}|$ with a duration of $|\mathcal{T}^\text{A}_i|$ time steps per activation. We repeat our 10-day optimization for the whole time span of the dataset, and our final results are the average based on these runs.

The overhead $M^\text{P}$ in \eqref{opt:preemp_times} per preemption is 1.5 minutes for AI-focused HPC data centers based on \cite{gu2019tiresias}, and 0.5 minutes for general-purpose HPC data centers, considering the avoidance of GPU communication compared to AI-focused HPC data centers. The extra workload due to preemption is limited below 1\% of the original computing workload, namely $\epsilon=1$\% in \eqref{opt:preemp_times}.

Our nominal parameters are set to $\overline{G}=1$ kW, $\overline{G_0}=0$, $\overline{A}=0.5$, $\overline{R}=1$ USD per hour per computing resource. The energy price $\overline{\pi}$ is set to the wholesale market price (front-of-the-meter) of 0.05 USD/kWh \cite{storage_joule}, considering the excessive power demand of hyperscale data centers. 

\begin{table*}[p]
\scriptsize
\centering
\caption{Summary of the analyzed HPC data centers}
\label{tab:summary}
\begin{tblr}{
  colspec = {
        Q[m,c,.07\linewidth]
        Q[m,c,.04\linewidth]
        Q[m,.27\linewidth]
        Q[m,c,.05\linewidth]
        Q[m,c,.075\linewidth]
        Q[m,.15\linewidth]
        Q[m,c,.07\linewidth]
        Q[m,c,.07\linewidth]
        },
  cell{5}{3} = {r=4}{},
  cell{5}{4} = {r=4}{},
  cell{5}{5} = {r=4}{},
  cell{5}{7} = {r=4}{},
  cell{5}{8} = {r=4}{},
  cell{10}{5} = {r=2}{},
  cell{10}{7} = {r=2}{},
  cell{10}{8} = {r=2}{},
  hline{1-2,16} = {-}{1.5pt},
  hline{3-15} = {-}{.25pt}
}
\textbf{Name}                    & \textbf{Type}            & \textbf{Main applications}                                                                                                                                                                                                                                                                                                                   & \textbf{Timespan}  & \textbf{Source}                                                                                & \textbf{System specs}                                                 & \textbf{Commercial} & \textbf{Academic}                                                       \\
Seren                                & AI                      & LLM (training/development), 7B to over 123B      \cite{seren}                                                                                                                                                                                                                                                                             & 03/2023 to 08/2023 & Shanghai AI Lab \cite{githubGitHubInternLMAcmeTrace}                          & 8*286 GPUs (A100), 128*286vCPUs, 1024*286 GB RAM                     & No                  & Yes       \\
Philly                               & AI                      & NN training only, CNN, RNN, LSTM \cite{jeon2019analysis}                                                                                                                                                                                                                                                                                                            & 08/2017 to 12/2017 & Microsoft's internal Philly clusters \cite{githubGitHubMsrfiddlephillytraces} & 2490 GPUs                                                            & Yes                 & No                                           \\
PAI                                  & AI                      & DL, RL, training and inference \cite{weng2022mlaas}                                                                                                                                                                                                                                                                                                               & 07/2020 to 08/2020 & Alibaba \cite{PAI_source}                                                                                        & 6742 GPUs (T4/P100/V100),  156576vCPUs                                & Yes                 & No               \\
Saturn                               & AI                      & DL for CV, NLP. and RL. various types of jobs in the DL development pipeline, e.g., data preprocessing, model training, inference, quantization, etc, but the majority is for training \cite{GPU_dataset}.                                                                                                                                                      & 04/2020 to 09/2020 & SenseTime \cite{githubGitHubSLabSystemGroupHeliosData}                        & 2096 GPUs (Pascal \& Volta), 64*262vCPUs, 256*262GB RAM & Yes                 & No                        \\
Earth                                &    AI              &                                                                                                                                                                                                                                                                                                                                              &                    &                                                                                                & 1144 GPUs (V100), 48*143vCPUs, 376*143GB RAM,~                        &                     &                                                                                           \\
Venus                                &      AI             &                                                                                                                                                                                                                                                                                                                                              &                    &                                                                                                & 1064 GPUs (V100), 48*133vCPUs, 376*133GB RAM,                         &                     &                                                                                        \\
Uranus                               &     AI              &                                                                                                                                                                                                                                                                                                                                              &                    &                                                                                                & 2112 GPUs (Pascal), 64*264vCPUs, 256*264GB RAM                         &                     &                                                                                      \\
ORNL                                 & General & General scientific computing, like Fluid Dynamics and physics accelerators (recorded in their data files) \cite{ornlLandingPage}                                                                                                                                                                                                                                 & 01/2015 to 12/2019 & ORNL \cite{ornlLandingPage}                                                   & 18688 nodes, 299,008 CPU cores, 18688 Kepler K20X GPUs \cite{ORNL_dataset}               & No                  & Yes                                   \\
MIRA                                 & General                      & General scientific computing, scientific research, including studies in the fields of material science, climatology, seismology, and computational chemistry \cite{mira_main_applications}.                                                                                                                                                                              & 04/2013 to 03/2020 & Argonne Leadership Computing Facility \cite{anlALCFPublic}                    & 786432 CPU cores                                                      & No                  & Yes                                        \\
THETA                                & General & Simulation, data analytics, AI \cite{anlThetaThetaGPUArgonne}, detailed molecular simulations to massive cosmological models \cite{hpcwireThetaSupercomputer}                                                                                                                                              & 07/2017 to 01/2024 & Argonne Leadership Computing Facility \cite{anlALCFPublic}                    & 281088 CPU cores for THETA + 24*128 (EPYC 7742) for THETA GPU         & No                  & Yes          \\
CIEMAT-Euler                         & General & Data science, AI applied to medicine, dark matter, proton collider analysis and Virgo gravitational waves identification, modelling and simulating of fluid mechanics: combustion, flame dynamics, acoustics \cite{ciemat2022}                                                                                              & 11/2008 to 12/2017 & Parallel Workloads Archive \cite{hujiParallelWorkloads}                       & 1920 CPU cores  (may include GPUs as the application includes AI, but not clearly stated)                      & No                  & Yes               \\
PM100                                & General & Computational Chemistry, Condensed Matter Physics and Computational Fluid Dynamics \cite{cinecaUserStatistics}, advanced fluid-dynamic simulations \cite{cinecaASPENINSTITUTE}, Nuclear Physics \cite{praceriPRACE22nd}, drug discovery \cite{hpcwireCINECAsMarconi100}. & 05/2020 to 10/2020 & Zenodo \cite{PM100_datasource}                                 & 980 nodes, 32 CPU cores, 4 V100 GPUs              & No                  & Yes             \\
University of Luxemburg Gaia Cluster & General                      & It is used mainly by biologists working with large data problems and engineering people working with physical simulations \cite{hujiParallelWorkloads}.                                                                                                                                                                                                                   & 05/2014 to 08/2014 & Parallel Workloads Archive \cite{hujiParallelWorkloads}                       & 2004 CPU cores                                                        & No                  & Yes           \\
MetaCentrum 2                        & General & Computer science, computational chemistry, biomedical computing, bioinformatics, physics, bioinformatics, computer science (middleware technologies), interoperability (research infrastructures) computer science (data platforms) \cite{docsAmberMetacentrum}.                                                            & 01/2013 to 12/2014 & Parallel Workloads Archive \cite{hujiParallelWorkloads}                       & 19 clusters, with 495 nodes and 8412 cores (include some GPU nodes)  & No                  & Yes                                 
\end{tblr}
\end{table*}

\newpage


\ifCLASSOPTIONcaptionsoff
  \newpage
\fi

\bibliographystyle{IEEEtran}
\bibliography{myref}
\vfill

\pagebreak \clearpage
\onecolumn
\setcounter{subsection}{0}
{\centering
\Huge
Supplementary File: AI-focused Data Centers Can Provide More Grid Flexibility and at Lower Cost

\vspace{3mm}
\ifdoubleblind
    \fontsize{11}{13.5}\selectfont
    Anonymous Authors\\
    Anonymous Affiliation\\
    \normalsize
\else
    \fontsize{11}{13.5}\selectfont
    Yihong Zhou\IEEEauthorrefmark{1},
    \'Angel Paredes\IEEEauthorrefmark{2},
    Chaimaa Essayeh\IEEEauthorrefmark{3},
    and Thomas Morstyn\IEEEauthorrefmark{4}
    \\
    \vspace{2mm}
    \IEEEauthorblockA{\IEEEauthorrefmark{1}School of Engineering, The University of Edinburgh, U.K., yihong.zhou@ed.ac.uk}\\
    \IEEEauthorblockA{\IEEEauthorrefmark{2}Department of Electrical Engineering, University of M\'alaga, Spain, angelparedes@uma.es} \\
    \IEEEauthorblockA{\IEEEauthorrefmark{3}Department of Engineering, Nottingham Trent University, U.K., chaimaa.essayeh@ntu.ac.uk}\\
    \IEEEauthorblockA{\IEEEauthorrefmark{4}Department of Engineering Science, University of Oxford, U.K., thomas.morstyn@eng.ox.ac.uk}\\
    \normalsize
\fi
}%

\begin{figure*}[ht!]
  \centering
  \subfloat[ ]{\includegraphics[width=.5\linewidth]{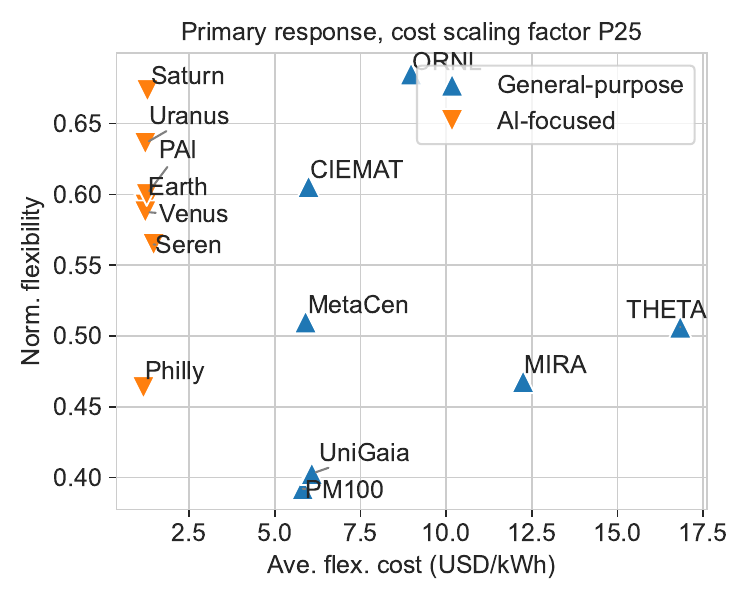}\label{fig:flex_vs_cost_dur0.25freq80_P25}}\hfill
  \subfloat[ ]{\includegraphics[width=.5\linewidth]{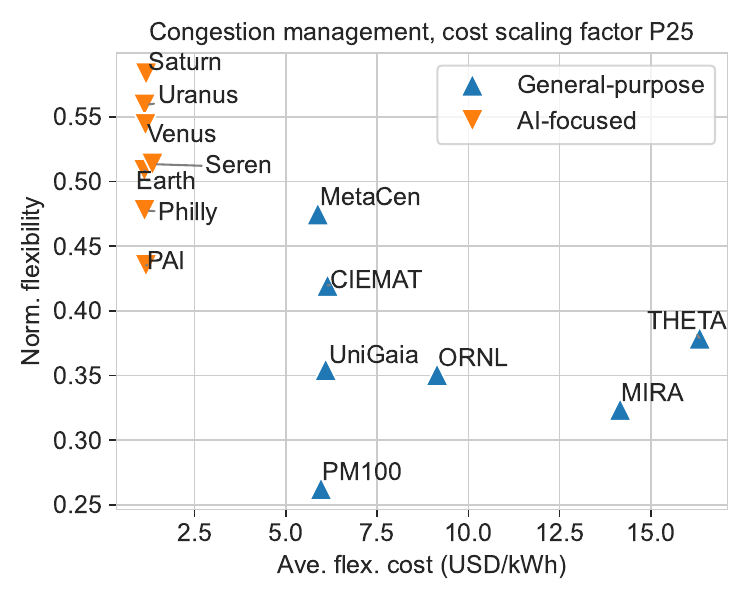}\label{fig:flex_vs_cost_dur2.0freq10_P25}} 

  \subfloat[ ]{\includegraphics[width=.5\linewidth]{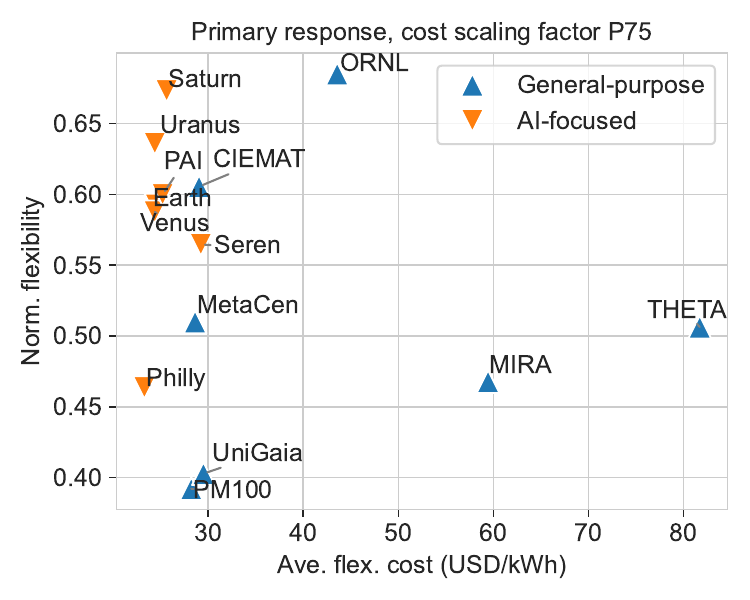}\label{fig:flex_vs_cost_dur0.25freq80_P75}}\hfill
  \subfloat[ ]{\includegraphics[width=.5\linewidth]{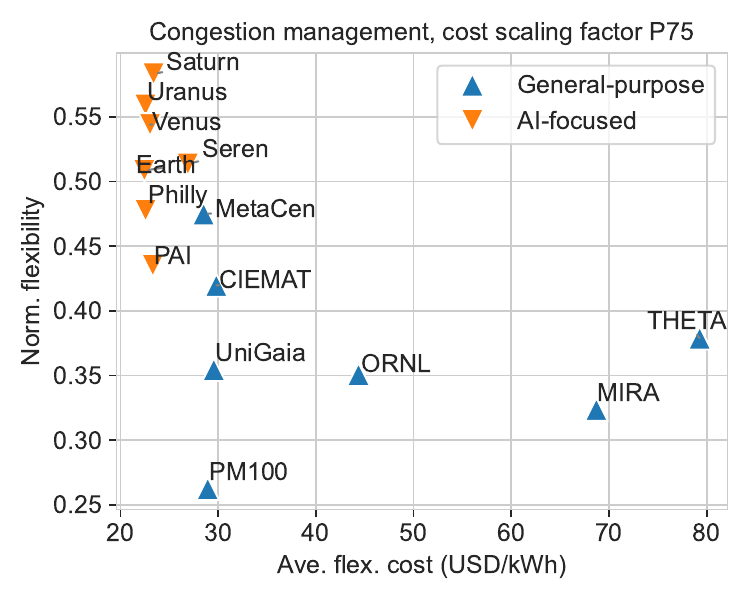}\label{fig:flex_vs_cost_dur2.0freq10_P75}} 

  \caption{The normalized maximum amount of flexibility (Norm. flexibility) versus the average flexibility cost when providing 100\% of the maximum flexibility (ave. flex. cost) for two power system services. \\\hspace{\textwidth} (a) Results for the primary response service under the 25th percentile (P25) of the cost scaling factor. (b) Results for the congestion management service under the 25th percentile (P25) of the cost scaling factor. (c) Results for the primary response service under the 75th percentile (P75) of the cost scaling factor. (d) Results for the congestion management service under the 75th percentile (P75) of the cost scaling factor. The cost scaling factor is estimated using data from Google Cloud, AWS, and Oracle. \\\hspace{\textwidth} 
  This figure leads to the same conclusion as Fig. 5 of our main manuscript.}
  \label{fig:all_HPC_analysis_appen}
  \vspace{-2 mm}
\end{figure*}

\begin{figure*}
  \centering
  \subfloat[ ]{\includegraphics[width=.5\linewidth]{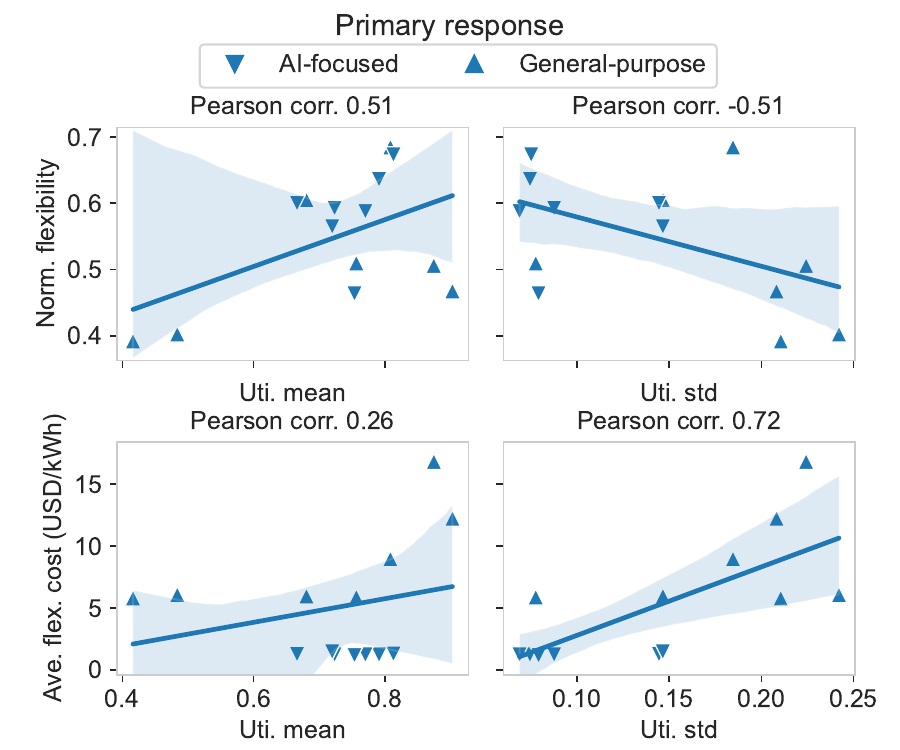}\label{fig:corrplot_dur0.25freq80_P25}}\hfill
  \subfloat[ ]{\includegraphics[width=.5\linewidth]{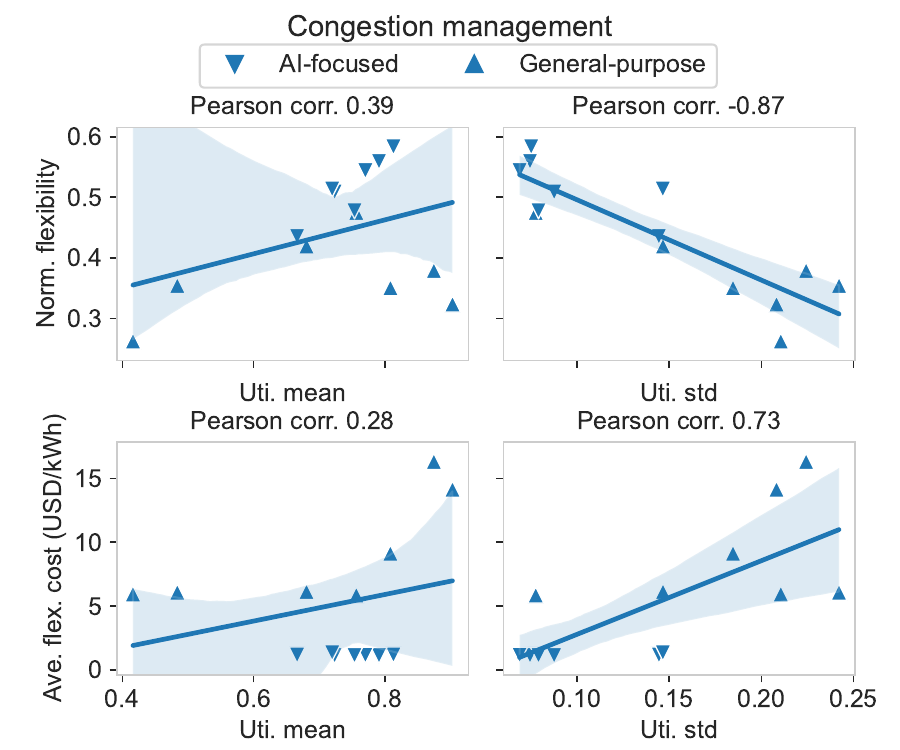}\label{fig:corrplot_dur2.0freq10_P25}}\hfill
    \subfloat[ ]{\includegraphics[width=.5\linewidth]{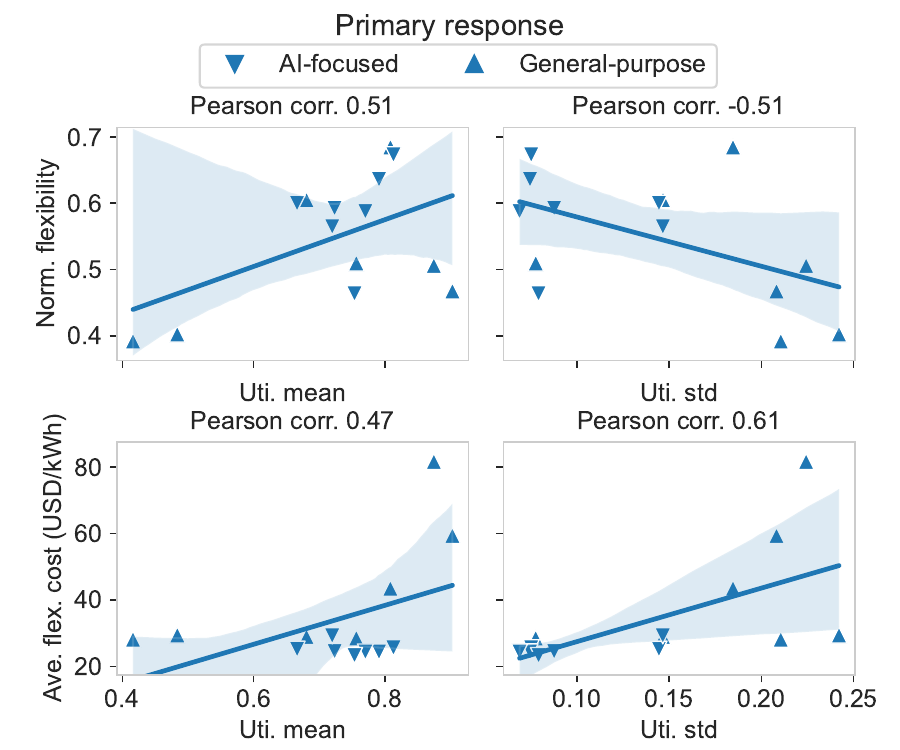}\label{fig:corrplot_dur0.25freq80_P75}}\hfill
  \subfloat[ ]{\includegraphics[width=.5\linewidth]{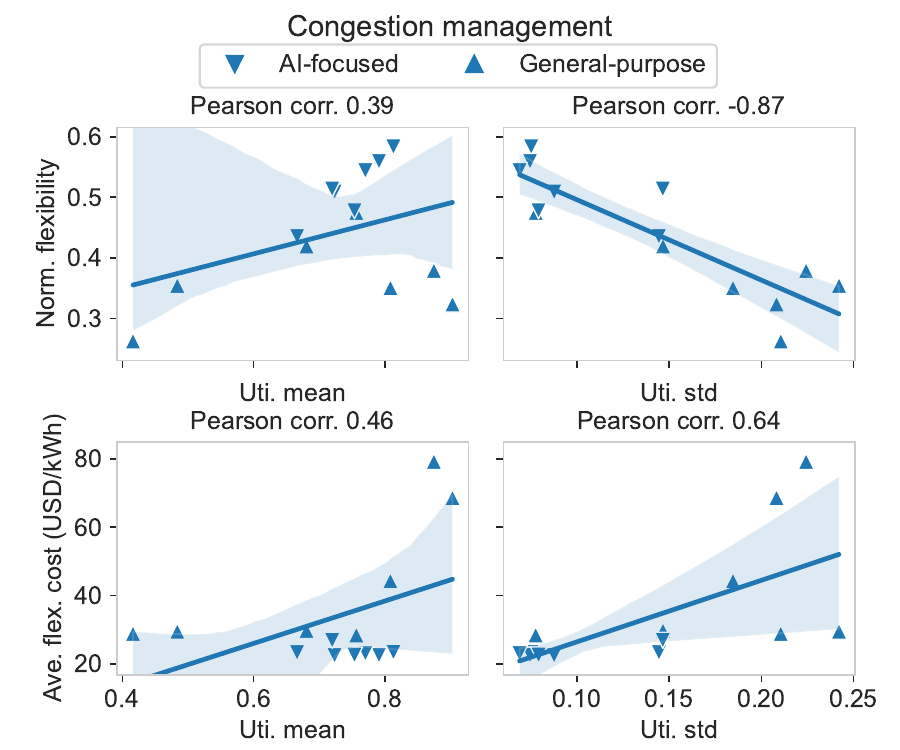}\label{fig:corrplot_dur2.0freq10_P75}}\hfill
  
   \vspace{-2 mm}
  \caption{Correlation analysis of the flexibility results and data center utilization patterns. (a) Correlation between the normalized maximum amount of flexibility (Norm. flexibility) for primary response, the average flexibility cost (Ave. flex. cost) for primary response, the mean utilization rate (Uti. mean), and the standard deviation of utilization (Uti. std). Regression lines with confidence intervals (shaded areas) are plotted for highlighting the trends. (b) Correlation results where the flexibility and cost are for congestion management. Plots (a) and (b) are for the 25th percentile of the cost scaling factor. Plots (c) and (d) are for the 75th percentile of the cost scaling factor. The cost scaling factor is estimated using data from Google Cloud, AWS, and Oracle.  \\\hspace{\textwidth}
  This figure leads to the same conclusion, as we have analyzed in Fig. 6 of our main manuscript.}
  \label{fig:all_HPC_corr_25_75}
\end{figure*}

\begin{figure*}
  \centering

  \subfloat[ ]{\includegraphics[width=.5\linewidth]{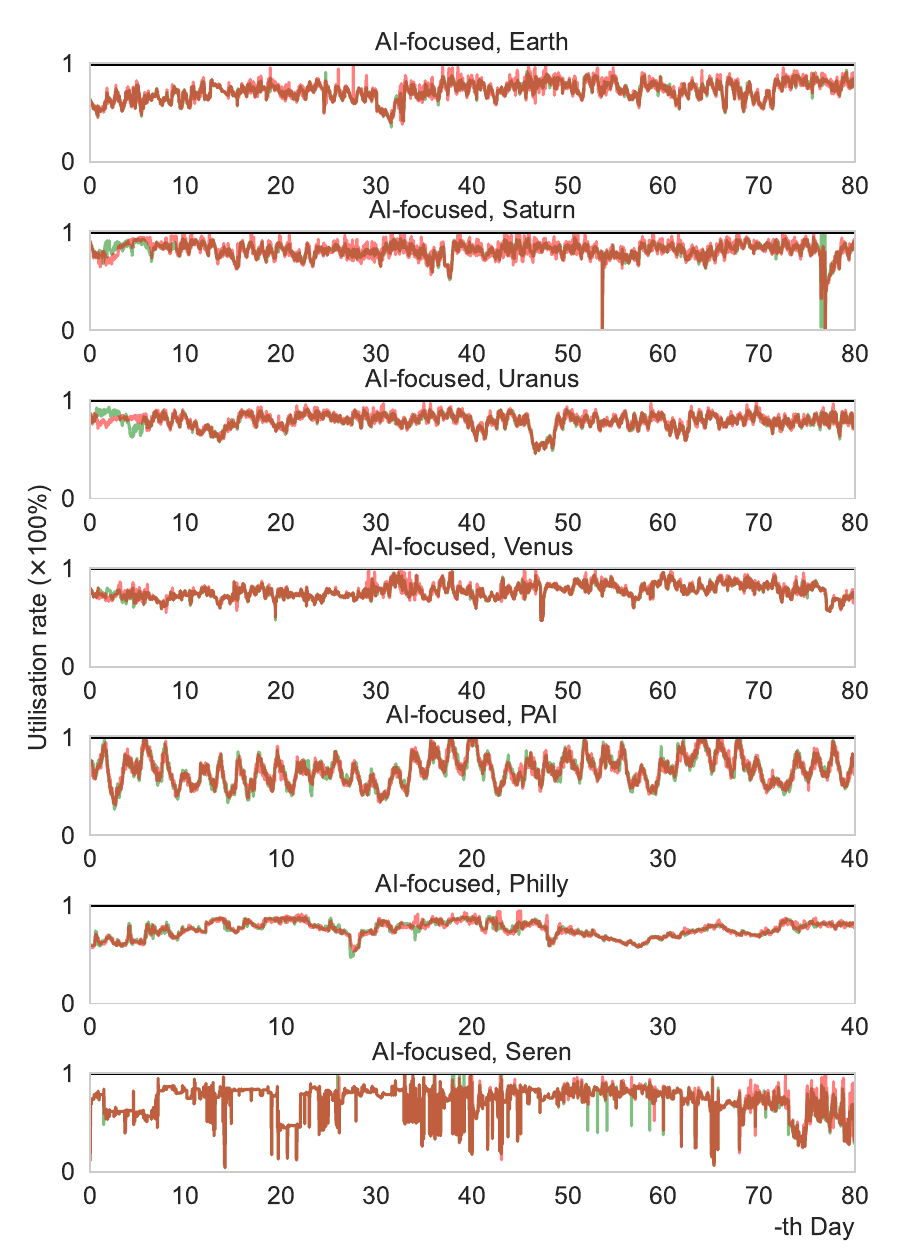}\label{fig:uti_base_AI_agg}}\hfill
  \subfloat[ ]{\includegraphics[width=.5\linewidth]{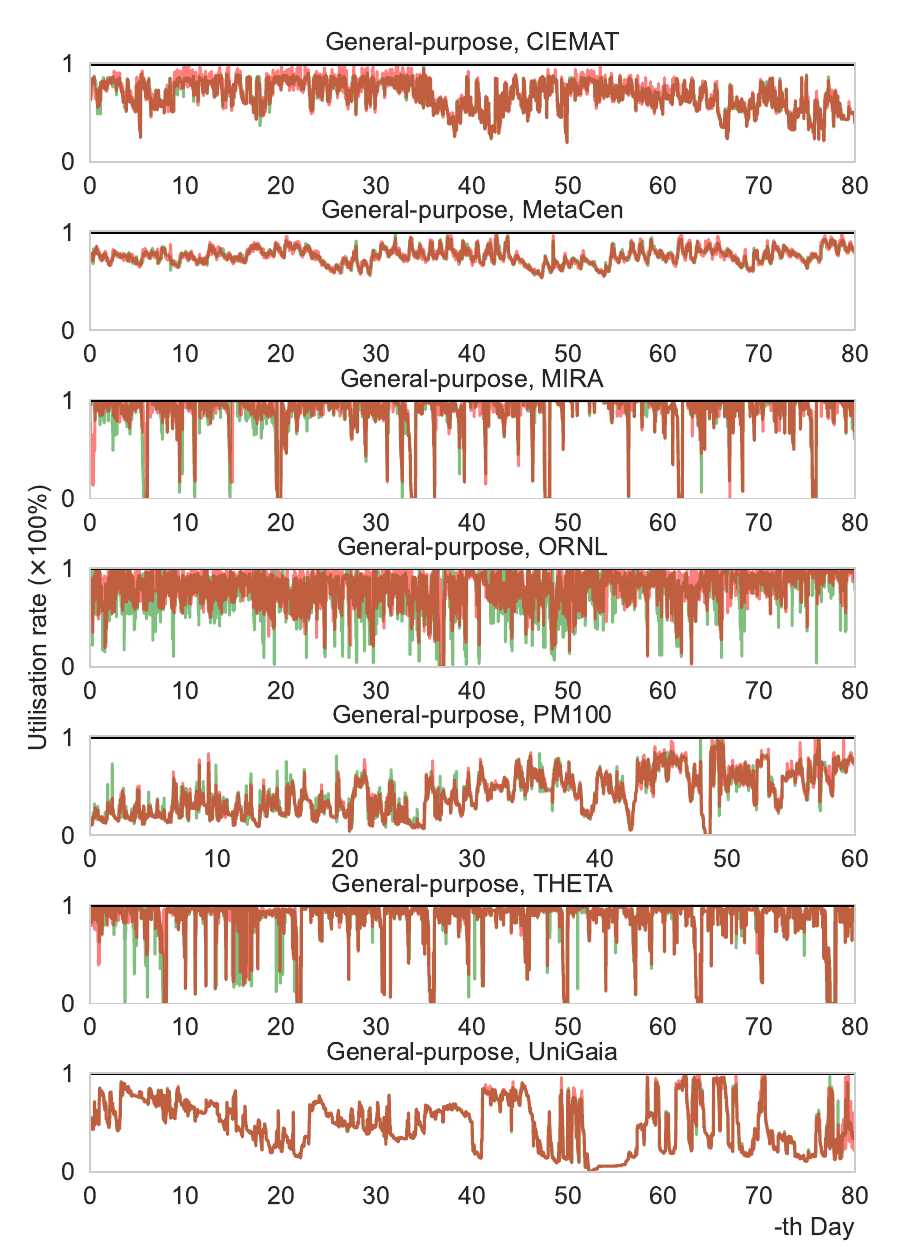}\label{fig:uti_base_General_agg}}
   \vspace{ -6mm}
  \caption{Baseline utilization time series of 14 data centers before and after job aggregation. (a) Utilization of the 7 AI-focused HPC data centers. (b) Utilization of the 7 general-purpose HPC data centers. The black lines refer to the utilization upper bound. The green lines are the utilization derived by the original job datasets. The red lines are the utilization derived by the aggregated jobs after job aggregation, which is a method used to reduce the computation complexity of solving optimization problems, as detailed in the Methods section. Is can be seen that the aggregated baseline profiles effectively capture the characteristics of the original profiles before aggregation (the green lines). Therefore, the applied aggregation strategy aims to improve computational efficiency while preserving utilization fidelity and potential flexibility. }
  \label{fig:util_base_all_HPCs_agg}
  \vspace{-2 mm}
\end{figure*}

\section*{Collecting Cloud Platform CPU and GPU pricing information}

\subsection{Computing prices and model specifications from cloud platforms}
\label{sec:criteria}

We collect GPU and CPU rental options available on Google Cloud, Amazon Web Services (AWS), Oracle, and Lambda Cloud. These cloud platforms provide price and machine specifications, where the latter enables us to link the price information to device power and computing speed from other sources. Here, we detail the source of each price and specification information. Note that on cloud platforms, they usually use the term ``instance'' to refer to our ``computing option''.
\begin{itemize}
    \item For Google Cloud, up-to-date price information and the GPU model can be found at their instance creation page, which requires a Google account log-in. The CPU model specification can be accessed here \cite{google_cpu_model}.
    \item For AWS, computing prices for all computing options can be found here \cite{aws_price}. Model specifications for both the GPU and CPU options are available here \cite{aws_model}.
    \item For Oracle, the computing price and model specifications are available here \cite{oracle_all_info}.
    \item Lambda Cloud provides computing prices and model specifications here \cite{lambda_gpu_cloud}. Note that, as mentioned in our main manuscript, GPU prices on Lambda Cloud are significantly lower than in other cloud platforms. Therefore, we analyzed the Lambda Cloud data separately as Figure 4.
\end{itemize}
Previous computing price information are for the whole machine. We divide the total price by the number of virtual CPUs (vCPUs) or GPUs to get the price per vCPU or per GPU. This can be justified by the observed proportional relationship between price and number of vCPUs/GPUs on cloud platforms. 

On most cloud platforms, CPUs are rented on the basis of vCPUs, which typically represent one thread of a physical CPU. We follow their convention here. Details on the rated power of both CPUs and GPUs are available on the respective manufacturers' websites. The power of a vCPU is calculated by dividing the rated power of the physical CPU by the number of vCPUs. 

It should be noted that cloud platforms provide CPUs not just for high-performance computing (HPC) but also for other purposes such as web services. As we focus on HPC data centers, we only collect a subset of CPU options on cloud platforms with optimized computing performance. The collection criteria are summarized below:
\begin{itemize}
    \item Google Cloud: For CPUs, we only collect the ``compute optimized'' instances, and several ``general'' instances with description ``consistently high performance''. Discounts are not applied. The price region is ``US-central-1''.
    \item AWS: We collect ``compute optimized'' instances and ``HPC optmized'' instances. We select ``on-demand'' pricing with Linux operation systems. The price zone is ``US East (N. Virginia)''.
    \item Oracle: Oracle does not have dedicated ``compute optimized'' instances. However, they recommend the ``bare metal'' type for HPC computing \cite{oracle_baremetal}. Therefore, we only select CPU options within that type.
    \item Lambda Cloud only provides GPU rental, so is not discussed here.
\end{itemize}


\subsection{Computing speed information}

PassMark CPU benchmark test \cite{passmark} provides multithread CPU ratings, which is used as our CPU speed. This rating is the weighted harmonic average of the computing speed for several benchmark tests, such as integer math, floating point math, physics, etc., that are important for HPC applications \cite{passmark_scoremean}.

GPU speed information is available at Lambda Lab \cite{lambda_gpu_benchmark}, which is the weighted average speed for a range of AI tasks. The speed is relative to the speed of a single Tesla V100 GPU, and is provided on both the FP32 and FP16 precision basis. We use the harmonic average of the two precisions as the final GPU speed. We use the 2023 benchmark speed results for up-to-dateness, unless only the 2022 speed information is available for a few GPU options. 



\subsection{Data file description}

In order to estimate the cost scaling factor (see Methods of the main manuscript), we collect price and speed information from cloud platforms and record them into ``all\_cloud\_data\_chart.xlsx'' file in the supplementary material. The file has the following columns:

\begin{enumerate}



    \item \textbf{Provider:} Specifies the cloud service provider (e.g., Google Cloud, AWS, Oracle Cloud).
    \item \textbf{Type:} Indicates whether the entry refers to a CPU (Central Processing Unit) or a GPU (Graphics Processing Unit).
    \item \textbf{Model:} The specific model of the CPU or GPU used in this computing option.
    \item \textbf{Number of vCPU / GPU:} Lists the number of vCPUs or GPUs in the computing option. An asterisk (*) indicates that the option is estimated instead of actually available on the cloud platform. For example, a Cloud platform may only rent 4 GPUs or 32 vCPUs as a whole. We can then estimate the price of 1 GPU or 1 vCPU, as long as the corresponding speed information is available. The price is estimated by, e.g., dividing the 4 GPU price by 4. This can be justified by the observed proportional relationship between the price and the number of vCPUs/GPUs on the cloud platforms. This is to enrich our dataset to create more accurate estimate; otherwise, there are only limited computing options which can only support limited estimation samples of the cost scaling factor.
    \item \textbf{Memory (RAM per CPU, VRAM per GPU):} Indicates the total RAM for CPUs or VRAM per GPU.
    \item \textbf{Unit Price (\$/(unit·h)):} The hourly cost in USD per vCPU or per GPU.
    \item \textbf{Total device price (\$/h):} The total hourly cost for the entire computing option, considering all vCPUs or GPUs.
    \item \textbf{CPU Score:} The CPU speed for this option, which is only used for general-purpose HPC data centers.
    \item \textbf{GPU FP32:} The FP32 GPU speed relative to a single ``Tesla V100", which is only used for AI-focused HPC data centers.
    \item \textbf{GPU FP16:} The FP16 GPU speed relative to a single ``Tesla V100", which is only used for AI-focused HPC data centers.
    \item \textbf{Unit Rated Power (W/(vCPU, GPU)):} The power consumption per vCPU or GPU, measured in watts (W).
    \item \textbf{Total Rated Power (W):} The total power consumption for all vCPUs or GPUs in the computing option.
    \item \textbf{Notes:} Provides additional information or clarifications about the data, such as assumptions or special conditions.
\end{enumerate}

\end{document}